\documentstyle[10pt,epsf,epsfig,oldlfont,pennames,hhline,multicol,array,
amstex,feynmf,dp_delphititle]{dp_delphi}
\makeindex
\def\DpPaperGroup{EP}
\def\DpPaperRef{2000-134}
\def\DpDate{31 October 2000}
\def\DpAuthors{DELPHI Collaboration}
\def\DpSubmit{(Euro. Phys. J. C19(2001)29)}
\def\DpTitle{{Search for sleptons in \\
{\mbox{$ {\mathrm e}^+ {\mathrm e}^- $}} collisions at \boldmath \\$\sqrt{s}
 = $183 to 189 GeV}}
\def\DpComment{ }
\def\DpEMail{ }

\begin{document}
\makeatletter
\newcount\@tempcntc
\def\@citex[#1]#2{\if@filesw\immediate\write\@auxout{\string\citation{#2}}\fi
  \@tempcnta\z@\@tempcntb\m@ne\def\@citea{}\@cite{\@for\@citeb:=#2\do
    {\@ifundefined
       {b@\@citeb}{\@citeo\@tempcntb\m@ne\@citea\def\@citea{,}{\bf ?}\@warning
       {Citation `\@citeb' on page \thepage \space undefined}}%
    {\setbox\z@\hbox{\global\@tempcntc0\csname b@\@citeb\endcsname\relax}%
     \ifnum\@tempcntc=\z@ \@citeo\@tempcntb\m@ne
       \@citea\def\@citea{,}\hbox{\csname b@\@citeb\endcsname}%
     \else
      \advance\@tempcntb\@ne
      \ifnum\@tempcntb=\@tempcntc
      \else\advance\@tempcntb\m@ne\@citeo
      \@tempcnta\@tempcntc\@tempcntb\@tempcntc\fi\fi}}\@citeo}{#1}}
\def\@citeo{\ifnum\@tempcnta>\@tempcntb\else\@citea\def\@citea{,}%
  \ifnum\@tempcnta=\@tempcntb\the\@tempcnta\else
   {\advance\@tempcnta\@ne\ifnum\@tempcnta=\@tempcntb \else \def\@citea{--}\fi
    \advance\@tempcnta\m@ne\the\@tempcnta\@citea\the\@tempcntb}\fi\fi}
 
\makeatother
\begin{titlepage}
\pagenumbering{roman}
\CERNpreprint{\DpPaperGroup}{\DpPaperRef} 
\date{{\small\DpDate}} 
\title{\DpTitle} 
\address{\DpAuthors} 
\begin{shortabs} 
\noindent
%
\newcommand{\GeVcc} {\mbox{$ {\mathrm{GeV}}/c^2 $}}
\newcommand{\GeV} {\mbox{$ {\mathrm{GeV}} $}}
\noindent

Data taken by the {\tt DELPHI} experiment at centre-of-mass energies of 
183~\GeV\ and 189~\GeV\ with a total integrated luminosity of 212 pb$^{-1}$ 
have been used to search for the supersymmetric partners of the electrons, 
muons, and taus in the context of the Minimal Supersymmetric Standard Model
(MSSM). The decay topologies searched for were the direct decay (${\mathrm 
{\tilde \ell}}\rightarrow\ell{\mathrm {\tilde \chi_1^0}}$), producing
acoplanar lepton pairs plus missing energy, and the cascade decay (${\mathrm
{\tilde \ell}} \rightarrow \ell{\mathrm {\tilde \chi_2^0}} \rightarrow
\ell\gamma{\mathrm   {\tilde \chi_1^0}}$), producing acoplanar lepton and
photon pairs plus missing energy. The observed number of events is in
agreement with Standard Model predictions. The 95$\%$ CL excluded mass limits
for selectrons, smuons and staus are \mbox{$m_{\tilde {e}} \le$ 
87~\GeVcc\:}, \mbox{$m_{\tilde {\mu}} \le$ 80~\GeVcc\:} and $m_{\tilde {\tau}} \le$
75~\GeVcc, respectively, for values of $\mu$=-200~\GeVcc\ and
tan$\beta$=1.5.  
\end{shortabs}
\vfill
\begin{center}
\DpSubmit \ \\ 
\DpComment \ \\
\DpEMail \ \\
\end{center}
\vfill
\clearpage
\headsep 10.0pt
\addtolength{\textheight}{10mm}
\addtolength{\footskip}{-5mm}
\begingroup
%
\newcommand{\DpName}[2]{\hbox{#1$^{\ref{#2}}$},\hfill}
\newcommand{\DpNameTwo}[3]{\hbox{#1$^{\ref{#2},\ref{#3}}$},\hfill}
\newcommand{\DpNameThree}[4]{\hbox{#1$^{\ref{#2},\ref{#3},\ref{#4}}$},\hfill}
\newskip\Bigfill \Bigfill = 0pt plus 1000fill
\newcommand{\DpNameLast}[2]{\hbox{#1$^{\ref{#2}}$}\hspace{\Bigfill}}
%
\footnotesize
\noindent
\DpName{P.Abreu}{LIP}
\DpName{W.Adam}{VIENNA}
\DpName{T.Adye}{RAL}
\DpName{P.Adzic}{DEMOKRITOS}
\DpName{I.Ajinenko}{SERPUKHOV}
\DpName{Z.Albrecht}{KARLSRUHE}
\DpName{T.Alderweireld}{AIM}
\DpName{G.D.Alekseev}{JINR}
\DpName{R.Alemany}{CERN}
\DpName{T.Allmendinger}{KARLSRUHE}
% ...
\normalsize
\endgroup
\titlefoot{Department of Physics and Astronomy, Iowa State
     University, Ames IA 50011-3160, USA
    \label{AMES}}
\titlefoot{Physics Department, Univ. Instelling Antwerpen,
     Universiteitsplein 1, B-2610 Antwerpen, Belgium \\
     \indent~~and IIHE, ULB-VUB,
     Pleinlaan 2, B-1050 Brussels, Belgium \\
     \indent~~and Facult\'e des Sciences,
     Univ. de l'Etat Mons, Av. Maistriau 19, B-7000 Mons, Belgium
    \label{AIM}}
\titlefoot{Physics Laboratory, University of Athens, Solonos Str.
     104, GR-10680 Athens, Greece
    \label{ATHENS}}
\titlefoot{Department of Physics, University of Bergen,
     All\'egaten 55, NO-5007 Bergen, Norway
    \label{BERGEN}}
\titlefoot{Dipartimento di Fisica, Universit\`a di Bologna and INFN,
     Via Irnerio 46, IT-40126 Bologna, Italy
    \label{BOLOGNA}}
\titlefoot{Centro Brasileiro de Pesquisas F\'{\i}sicas, rua Xavier Sigaud 150,
     BR-22290 Rio de Janeiro, Brazil \\
     \indent~~and Depto. de F\'{\i}sica, Pont. Univ. Cat\'olica,
     C.P. 38071 BR-22453 Rio de Janeiro, Brazil \\
     \indent~~and Inst. de F\'{\i}sica, Univ. Estadual do Rio de Janeiro,
     rua S\~{a}o Francisco Xavier 524, Rio de Janeiro, Brazil
    \label{BRASIL}}
\titlefoot{Comenius University, Faculty of Mathematics and Physics,
     Mlynska Dolina, SK-84215 Bratislava, Slovakia
    \label{BRATISLAVA}}
\titlefoot{Coll\`ege de France, Lab. de Physique Corpusculaire, IN2P3-CNRS,
     FR-75231 Paris Cedex 05, France
    \label{CDF}}
\titlefoot{CERN, CH-1211 Geneva 23, Switzerland
    \label{CERN}}
\titlefoot{Institut de Recherches Subatomiques, IN2P3 - CNRS/ULP - BP20,
     FR-67037 Strasbourg Cedex, France
    \label{CRN}}
\titlefoot{Now at DESY-Zeuthen, Platanenallee 6, D-15735 Zeuthen, Germany
    \label{DESY}}
\titlefoot{Institute of Nuclear Physics, N.C.S.R. Demokritos,
     P.O. Box 60228, GR-15310 Athens, Greece
    \label{DEMOKRITOS}}
\titlefoot{FZU, Inst. of Phys. of the C.A.S. High Energy Physics Division,
     Na Slovance 2, CZ-180 40, Praha 8, Czech Republic
    \label{FZU}}
\titlefoot{Dipartimento di Fisica, Universit\`a di Genova and INFN,
     Via Dodecaneso 33, IT-16146 Genova, Italy
    \label{GENOVA}}
\titlefoot{Institut des Sciences Nucl\'eaires, IN2P3-CNRS, Universit\'e
     de Grenoble 1, FR-38026 Grenoble Cedex, France
    \label{GRENOBLE}}
\titlefoot{Helsinki Institute of Physics, HIP,
     P.O. Box 9, FI-00014 Helsinki, Finland
    \label{HELSINKI}}
\titlefoot{Joint Institute for Nuclear Research, Dubna, Head Post
     Office, P.O. Box 79, RU-101 000 Moscow, Russian Federation
    \label{JINR}}
\titlefoot{Institut f\"ur Experimentelle Kernphysik,
     Universit\"at Karlsruhe, Postfach 6980, DE-76128 Karlsruhe,
     Germany
    \label{KARLSRUHE}}
\titlefoot{Institute of Nuclear Physics and University of Mining and Metalurgy,
     Ul. Kawiory 26a, PL-30055 Krakow, Poland
    \label{KRAKOW}}
\titlefoot{Universit\'e de Paris-Sud, Lab. de l'Acc\'el\'erateur
     Lin\'eaire, IN2P3-CNRS, B\^{a}t. 200, FR-91405 Orsay Cedex, France
    \label{LAL}}
\titlefoot{School of Physics and Chemistry, University of Lancaster,
     Lancaster LA1 4YB, UK
    \label{LANCASTER}}
\titlefoot{LIP, IST, FCUL - Av. Elias Garcia, 14-$1^{o}$,
     PT-1000 Lisboa Codex, Portugal
    \label{LIP}}
\titlefoot{Department of Physics, University of Liverpool, P.O.
     Box 147, Liverpool L69 3BX, UK
    \label{LIVERPOOL}}
\titlefoot{LPNHE, IN2P3-CNRS, Univ.~Paris VI et VII, Tour 33 (RdC),
     4 place Jussieu, FR-75252 Paris Cedex 05, France
    \label{LPNHE}}
\titlefoot{Department of Physics, University of Lund,
     S\"olvegatan 14, SE-223 63 Lund, Sweden
    \label{LUND}}
\titlefoot{Universit\'e Claude Bernard de Lyon, IPNL, IN2P3-CNRS,
     FR-69622 Villeurbanne Cedex, France
    \label{LYON}}
\titlefoot{Univ. d'Aix - Marseille II - CPP, IN2P3-CNRS,
     FR-13288 Marseille Cedex 09, France
    \label{MARSEILLE}}
\titlefoot{Dipartimento di Fisica, Universit\`a di Milano and INFN-MILANO,
     Via Celoria 16, IT-20133 Milan, Italy
    \label{MILANO}}
\titlefoot{Dipartimento di Fisica, Univ. di Milano-Bicocca and
     INFN-MILANO, Piazza delle Scienze 2, IT-20126 Milan, Italy
    \label{MILANO2}}
\titlefoot{Niels Bohr Institute, Blegdamsvej 17,
     DK-2100 Copenhagen {\O}, Denmark
    \label{NBI}}
\titlefoot{IPNP of MFF, Charles Univ., Areal MFF,
     V Holesovickach 2, CZ-180 00, Praha 8, Czech Republic
    \label{NC}}
\titlefoot{NIKHEF, Postbus 41882, NL-1009 DB
     Amsterdam, The Netherlands
    \label{NIKHEF}}
\titlefoot{National Technical University, Physics Department,
     Zografou Campus, GR-15773 Athens, Greece
    \label{NTU-ATHENS}}
\titlefoot{Physics Department, University of Oslo, Blindern,
     NO-1000 Oslo 3, Norway
    \label{OSLO}}
\titlefoot{Dpto. Fisica, Univ. Oviedo, Avda. Calvo Sotelo
     s/n, ES-33007 Oviedo, Spain
    \label{OVIEDO}}
\titlefoot{Department of Physics, University of Oxford,
     Keble Road, Oxford OX1 3RH, UK
    \label{OXFORD}}
\titlefoot{Dipartimento di Fisica, Universit\`a di Padova and
     INFN, Via Marzolo 8, IT-35131 Padua, Italy
    \label{PADOVA}}
\titlefoot{Rutherford Appleton Laboratory, Chilton, Didcot
     OX11 OQX, UK
    \label{RAL}}
\titlefoot{Dipartimento di Fisica, Universit\`a di Roma II and
     INFN, Tor Vergata, IT-00173 Rome, Italy
    \label{ROMA2}}
\titlefoot{Dipartimento di Fisica, Universit\`a di Roma III and
     INFN, Via della Vasca Navale 84, IT-00146 Rome, Italy
    \label{ROMA3}}
\titlefoot{DAPNIA/Service de Physique des Particules,
     CEA-Saclay, FR-91191 Gif-sur-Yvette Cedex, France
    \label{SACLAY}}
\titlefoot{Instituto de Fisica de Cantabria (CSIC-UC), Avda.
     los Castros s/n, ES-39006 Santander, Spain
    \label{SANTANDER}}
\titlefoot{Dipartimento di Fisica, Universit\`a degli Studi di Roma
     La Sapienza, Piazzale Aldo Moro 2, IT-00185 Rome, Italy
    \label{SAPIENZA}}
\titlefoot{Inst. for High Energy Physics, Serpukov
     P.O. Box 35, Protvino, (Moscow Region), Russian Federation
    \label{SERPUKHOV}}
\titlefoot{J. Stefan Institute, Jamova 39, SI-1000 Ljubljana, Slovenia
     and Laboratory for Astroparticle Physics,\\
     \indent~~Nova Gorica Polytechnic, Kostanjeviska 16a, SI-5000 Nova Gorica, Slovenia, \\
     \indent~~and Department of Physics, University of Ljubljana,
     SI-1000 Ljubljana, Slovenia
    \label{SLOVENIJA}}
\titlefoot{Fysikum, Stockholm University,
     Box 6730, SE-113 85 Stockholm, Sweden
    \label{STOCKHOLM}}
\titlefoot{Dipartimento di Fisica Sperimentale, Universit\`a di
     Torino and INFN, Via P. Giuria 1, IT-10125 Turin, Italy
    \label{TORINO}}
\titlefoot{Dipartimento di Fisica, Universit\`a di Trieste and
     INFN, Via A. Valerio 2, IT-34127 Trieste, Italy \\
     \indent~~and Istituto di Fisica, Universit\`a di Udine,
     IT-33100 Udine, Italy
    \label{TU}}
\titlefoot{Univ. Federal do Rio de Janeiro, C.P. 68528
     Cidade Univ., Ilha do Fund\~ao
     BR-21945-970 Rio de Janeiro, Brazil
    \label{UFRJ}}
\titlefoot{Department of Radiation Sciences, University of
     Uppsala, P.O. Box 535, SE-751 21 Uppsala, Sweden
    \label{UPPSALA}}
\titlefoot{IFIC, Valencia-CSIC, and D.F.A.M.N., U. de Valencia,
     Avda. Dr. Moliner 50, ES-46100 Burjassot (Valencia), Spain
    \label{VALENCIA}}
\titlefoot{Institut f\"ur Hochenergiephysik, \"Osterr. Akad.
     d. Wissensch., Nikolsdorfergasse 18, AT-1050 Vienna, Austria
    \label{VIENNA}}
\titlefoot{Inst. Nuclear Studies and University of Warsaw, Ul.
     Hoza 69, PL-00681 Warsaw, Poland
    \label{WARSZAWA}}
\titlefoot{Fachbereich Physik, University of Wuppertal, Postfach
     100 127, DE-42097 Wuppertal, Germany
    \label{WUPPERTAL}}
\addtolength{\textheight}{-10mm}
\addtolength{\footskip}{5mm}
\clearpage
\headsep 30.0pt
\end{titlepage}
%
\pagenumbering{arabic} 
\setcounter{footnote}{0} %
\large

\def\leqsim{\mathbin{\;\raise1pt\hbox{$<$}\kern-8pt\lower3pt\hbox{$\sim$}\;}}
\def\geqsim{\mathbin{\;\raise1pt\hbox{$>$}\kern-8pt\lower3pt\hbox{$\sim$}\;}}
\renewcommand{\dfrac}[2]{\frac{\displaystyle #1}{\displaystyle #2}}
\renewcommand\topfraction{1.}
\renewcommand\bottomfraction{1.}
\renewcommand\floatpagefraction{0.}
\renewcommand\textfraction{0.}
\def\MXN#1{\mbox{$ m_{\tilde{\chi}^0_#1} $}}
\def\MXC#1{\mbox{$ M_{\tilde{\chi}^{\pm}_#1} $}}
\def\XP#1{\mbox{$ \tilde{\chi}^+_#1 $}}
\def\XM#1{\mbox{$ \tilde{\chi}^-_#1 $}}
\def\XPM#1{\mbox{$ \tilde{\chi}^{\pm}_#1 $}}
\def\XN#1{\mbox{$ \tilde{\chi}^0_#1 $}}
\newcommand{\tanb} {\mbox{$ \tan \beta $}}
\newcommand{\smu} {\mbox{$ \tilde{\mu} $}}
\newcommand{\smur} {\mbox{$ \tilde{\mathrm \mu}_R $}}
\newcommand{\msmu} {\mbox{$ M_{\tilde{\mu}} $}}
\newcommand{\msmur} {\mbox{$ M_{\tilde{\mu}_R} $}}
\newcommand{\msmul} {\mbox{$ M_{\tilde{\mu}_L} $}}
\newcommand{\sel} {\mbox{$ \tilde{\mathrm e} $}}
\newcommand{\selr} {\mbox{$ \tilde{\mathrm e}_R $}}
\newcommand{\msel} {\mbox{$ M_{\tilde{\mathrm e}} $}}
\newcommand{\msell} {\mbox{$ M_{\tilde{\mathrm e}_L} $}}
\newcommand{\mselr} {\mbox{$ M_{\tilde{\mathrm e}_R} $}}
\newcommand{\sta} {\mbox{$ \tilde{\tau} $}}
\newcommand{\staur} {\mbox{$ \tilde{\mathrm \tau}_R $}}
\newcommand{\staum} {\mbox{$ \tilde{\mathrm \tau}_{min} $}}
\newcommand{\msta} {\mbox{$ m_{\tilde{\tau}} $}}
\newcommand{\mstal} {\mbox{$ M_{\tilde{\tau}_L} $}}
\newcommand{\mstar} {\mbox{$ M_{\tilde{\tau}_R} $}}
\newcommand{\sfe} {\mbox{$ \tilde{\mathrm f} $}}
\newcommand{\msfe} {\mbox{$ M_{\tilde{\mathrm f}} $}}
\newcommand{\sle} {\mbox{$ \tilde{\ell} $}}
\newcommand{\msle} {\mbox{$ M_{\tilde{\ell}} $}}
\newcommand{\stq} {\mbox{$ \tilde {\mathrm t} $}}
\newcommand{\mstq} {\mbox{$ M_{\tilde {\mathrm t}} $}}
\newcommand{\sbq} {\mbox{$ \tilde {\mathrm b} $}}
\newcommand{\msbq} {\mbox{$ M_{\tilde {\mathrm b}} $}}
\newcommand{\An} {\mbox{$ {\mathrm A}^0 $}}
\newcommand{\hn} {\mbox{$ {\mathrm h}^0 $}}
\newcommand{\Zn} {\mbox{$ {\mathrm Z}^0 $}}
\newcommand{\Hn} {\mbox{$ {\mathrm H}^0 $}}
\newcommand{\HP} {\mbox{$ {\mathrm H}^+ $}}
\newcommand{\HM} {\mbox{$ {\mathrm H}^- $}}
\newcommand{\Wp} {\mbox{$ {\mathrm W}^+ $}}
\newcommand{\Wm} {\mbox{$ {\mathrm W}^- $}}
\newcommand{\WW} {\mbox{$ {\mathrm W}^+{\mathrm W}^- $}}
\newcommand{\ZZ} {\mbox{$ {\mathrm Z}^0{\mathrm Z}^0 $}}
\newcommand{\HZ} {\mbox{$ {\mathrm H}^0 {\mathrm Z}^0 $}}
\newcommand{\GW} {\mbox{$ \Gamma_{\mathrm W} $}}
\newcommand{\Zg} {\mbox{$ \Zn \gamma $}}
\newcommand{\sqs} {\mbox{$ \sqrt{s} $}}
\newcommand{\epm} {\mbox{$ {\mathrm e}^{\pm} $}}
\newcommand{\ee} {\mbox{$ {\mathrm e}^+ {\mathrm e}^- $}}
\newcommand{\eeto} {\mbox{$ {\mathrm e}^+ {\mathrm e}^- \to $}}
\newcommand{\ellell} {\mbox{$ \ell^+ \ell^- $}}
\newcommand{\eeWW} {\mbox{$ \ee \rightarrow \WW $}}
\newcommand{\MeV} {\mbox{$ {\mathrm{MeV}} $}}
\newcommand{\MeVc} {\mbox{$ {\mathrm{MeV}}/c $}}
\newcommand{\MeVcc} {\mbox{$ {\mathrm{MeV}}/c^2 $}}
\newcommand{\GeV} {\mbox{$ {\mathrm{GeV}} $}}
\newcommand{\GeVc} {\mbox{$ {\mathrm{GeV}}/c $}}
\newcommand{\GeVcc} {\mbox{$ {\mathrm{GeV}}/c^2 $}}
\newcommand{\TeV} {\mbox{$ {\mathrm{TeV}} $}}
\newcommand{\TeVc} {\mbox{$ {\mathrm{TeV}}/c $}}
\newcommand{\TeVcc} {\mbox{$ {\mathrm{TeV}}/c^2 $}}
\newcommand{\MZ} {\mbox{$ m_{{\mathrm Z}^0} $}}
\newcommand{\MW} {\mbox{$ m_{\mathrm W} $}}
\newcommand{\MA} {\mbox{$ m_{\mathrm A} $}}
\newcommand{\GF} {\mbox{$ {\mathrm G}_{\mathrm F} $}}
\newcommand{\MH} {\mbox{$ m_{{\mathrm H}^0} $}}
\newcommand{\MHP} {\mbox{$ m_{{\mathrm H}^\pm} $}}
\newcommand{\MSH} {\mbox{$ m_{{\mathrm h}^0} $}}
\newcommand{\MT} {\mbox{$ m_{\mathrm t} $}}
\newcommand{\GZ} {\mbox{$ \Gamma_{{\mathrm Z}^0} $}}
\renewcommand{\SS} {\mbox{$ \mathrm S $}}
\newcommand{\TT} {\mbox{$ \mathrm T $}}
\newcommand{\UU} {\mbox{$ \mathrm U $}}
\newcommand{\alphmz} {\mbox{$ \alpha (m_{{\mathrm Z}^0}) $}}
\newcommand{\alphas} {\mbox{$ \alpha_{\mathrm s} $}}
\newcommand{\alphmsb} {\mbox{$ \alphas (m_{\mathrm Z})
_{\overline{\mathrm{MS}}} $}}
\newcommand{\alphbar} {\mbox{$ \overline{\alpha}_{\mathrm s} $}}
\newcommand{\Ptau} {\mbox{$ P_{\tau} $}}
\newcommand{\mean}[1] {\mbox{$ \left\langle #1 \right\rangle $}}
\newcommand{\dgree} {\mbox{$ ^\circ $}}
\newcommand{\qqg} {\mbox{$ {\mathrm q}\bar{\mathrm q}\gamma $}}
\newcommand{\Wev} {\mbox{$ {\mathrm{W e}} \nu_{\mathrm e} $}}
\newcommand{\Zvv} {\mbox{$ \Zn \nu \bar{\nu} $}}
\newcommand{\Zee} {\mbox{$ \Zn \ee $}}
\newcommand{\ctw} {\mbox{$ \cos\theta_{\mathrm W} $}}
\newcommand{\thw} {\mbox{$ \theta_{\mathrm W} $}}
\newcommand{\thetabar}{\mbox{$ \theta^* $}}
\newcommand{\phibar} {\mbox{$ \phi^* $}}
\newcommand{\thetapl} {\mbox{$ \theta_+ $}}
\newcommand{\phipl} {\mbox{$ \phi_+ $}}
\newcommand{\thetamin}{\mbox{$ \theta_- $}}
\newcommand{\phimin} {\mbox{$ \phi_- $}}
\newcommand{\ds} {\mbox{$ {\mathrm d} \sigma $}}
\newcommand{\jjlv} {\mbox{$ j j \ell \nu $}}
\newcommand{\jjjj} {\mbox{$ j j j j $}}
\newcommand{\jjvv} {\mbox{$ j j \nu \bar{\nu} $}}
\newcommand{\qqvv} {\mbox{$ \mathrm{q \bar{q}} \nu \bar{\nu} $}}
\newcommand{\qqll} {\mbox{$ \mathrm{q \bar{q}} \ell \bar{\ell} $}}
\newcommand{\jjll} {\mbox{$ j j \ell \bar{\ell} $}}
\newcommand{\lvlv} {\mbox{$ \ell \nu \ell \nu $}}
\newcommand{\dz} {\mbox{$ \delta g_{\mathrm{W W Z} } $}}
\newcommand{\pT} {\mbox{$ p_{\mathrm{T}} $}}
\newcommand{\ptr} {\mbox{$ p_{\perp} $}}
\newcommand{\ptrjet} {\mbox{$ p_{\perp {\mathrm{jet}}} $}}
\newcommand{\Wvis} {\mbox{$ {\mathrm W}_{\mathrm{vis}} $}}
\newcommand{\gamgam} {\mbox{$ \gamma \gamma $}}
\newcommand{\qaqb} {\mbox{$ {\mathrm q}_1 \bar{\mathrm q}_2 $}}
\newcommand{\qcqd} {\mbox{$ {\mathrm q}_3 \bar{\mathrm q}_4 $}}
\newcommand{\bbbar} {\mbox{$ {\mathrm b}\bar{\mathrm b} $}}
\newcommand{\ffbar} {\mbox{$ {\mathrm f}\bar{\mathrm f} $}}
\newcommand{\qqbar} {\mbox{$ {\mathrm q}\bar{\mathrm q} $}}
\newcommand{\nunubar} {\mbox{$ {\nu}\bar{\nu} $}}
\newcommand{\qqbarp} {\mbox{$ {\mathrm q'}\bar{\mathrm q}' $}}
\newcommand{\djoin} {\mbox{$ d_{\mathrm{join}} $}}
\newcommand{\mErad} {\mbox{$ \left\langle E_{\mathrm{rad}} \right\rangle $}}
\newcommand{\Lum}{${\cal L}\;$}
\newcommand{\lum}{{\cal L}}
\newcommand{\Cms}{$\mbox{ cm}^{-2} \mbox{ s}^{-1}\;$}
\newcommand{\cms}{\mbox{ cm}^{-2} \mbox{ s}^{-1}\;}
\newcommand{\Ecms} {\mbox{$ E_{\mathrm{cms}} $}}
\newcommand{\Evis} {\mbox{$ E_{\mathrm{vis}} $}}
\newcommand{\Erad} {\mbox{$ E_{\mathrm{rad}} $}}
\newcommand{\Mvis} {\mbox{$ M_{\mathrm{vis}} $}}
\newcommand{\pvis} {\mbox{$ p_{\mathrm{vis}} $}}
\newcommand{\Minv} {\mbox{$ M_{\mathrm{inv}} $}}
\newcommand{\Mhfit}{\; \hat{m}_{H^0} }
\newcommand{\bl} {\mbox{\ \ \ \ \ \ \ \ \ \ } }
\newcommand{\Zto} {\mbox{$\mathrm Z^0 \to$}}
\newcommand{\dm} {\mbox{$\Delta M$}}
\newcommand{\lsp}{\relax\ifmmode{\mathrm{\widetilde{\chi}^0_1}}\else${\mathrm{\widetilde{\chi}^0_1}}$\fi}
\newcommand{\nlsp}{\relax\ifmmode{\mathrm{\widetilde{\chi}^0_2}}\else${\mathrm{\widetilde{\chi}^0_2}}$\fi}
\def\NPB#1#2#3{{\rm Nucl.~Phys.} {\bf{B#1}} (19#2) #3}
\def\PLB#1#2#3{{\rm Phys.~Lett.} {\bf{B#1}} (19#2) #3}
\def\PRD#1#2#3{{\rm Phys.~Rev.} {\bf{D#1}} (19#2) #3}
\def\PRL#1#2#3{{\rm Phys.~Rev.~Lett.} {\bf{#1}} (19#2) #3}
\def\ZPC#1#2#3{{\rm Z.~Phys.} {\bf C#1} (19#2) #3}
\def\PTP#1#2#3{{\rm Prog.~Theor.~Phys.} {\bf#1} (19#2) #3}
\def\MPL#1#2#3{{\rm Mod.~Phys.~Lett.} {\bf#1} (19#2) #3}
\def\PR#1#2#3{{\rm Phys.~Rep.} {\bf#1} (19#2) #3}
\def\RMP#1#2#3{{\rm Rev.~Mod.~Phys.} {\bf#1} (19#2) #3}
\def\HPA#1#2#3{{\rm Helv.~Phys.~Acta} {\bf#1} (19#2) #3}
\def\NIMA#1#2#3{{\rm Nucl.~Instr.~and~Meth.} {\bf{A#1}} (19#2) #3}
\def\CPC#1#2#3{{\rm Comp.~Phys.~Comm.} {\bf#1} (19#2) #3}
\newcommand{\etal}{{\it et al.}}
%
%
\def\ep{\mbox{$\mathrm{e}^{+}$}}
\def\em{\mbox{$\mathrm{e}^{-}$}}
\newcommand{\stql} {\mbox{$ {\mathrm \tilde{t}_{L}} $}}
\newcommand{\stqr} {\mbox{$ {\mathrm \tilde{t}_{R}} $}}
\newcommand{\sbql} {\mbox{$ {\mathrm \tilde{b}_{L}} $}}
\newcommand{\sbqr} {\mbox{$ {\mathrm \tilde{b}_{R}} $}}
\newcommand{\gvino} {\mbox{$ \tilde {\mathrm G} $}}
%
\newcommand{\chiz}[2]{\mbox{$\tilde{\chi}^{\mbox{#1}}_{\mbox{#2}}$}}
\newcommand {\neutralino} {\tilde{\chi }^{0}_{1}}
\newcommand {\neutrala} {\tilde{\chi }^{0}_{2}}

%
\section{Introduction \label{sec:INTRO}}
\setcounter{page}{1}
During the 1997 and 1998 data taking period, the LEP accelerator operated
at centre-of-mass energies of 183~GeV and 189~GeV respectively.
This allowed an extension of the searches for scalar
partners of electrons, muons, and taus, predicted by supersymmetric models,
over the limits on the 
production of these particles obtained from data previously taken
at centre-of-mass energies of 130-172~\GeV~\cite{183PUB}.  This 
paper reports on a search for these particles using the 212 pb$^{-1}$ of data 
taken by {\tt DELPHI} during 1997 and 1998. Similar searches have been
performed by other collaborations~\cite{LEP183}.

\par 

For a realistic experimental search one has to make some well motivated
assumptions. In this analysis, the model assumed is the Minimal
Supersymmetric Standard Model (MSSM)~\cite{MSSM}.  In the case that the MSSM
is locally invariant (often referred to as {\it minimal supergravity}), the
number of free parameters set at the unification scale (the scale at
which gauge couplings unify) can be reduced to five\footnote{See
  ~\cite{HERBI} ~\cite{PRIMER}~\cite{DREES} and references therein for
  further information on the actual supersymmetry breaking mechanism, and
  motivation for the assumptions made.}:   
\begin{center} $\tilde{m}_{0}$,\hspace{0.2cm}
  ${\tilde{m}}_{1/2},\hspace{0.2cm}A,\hspace{0.2cm}  B,\hspace{0.2cm}
   \mu \hspace{0.04cm} .$
\end{center}
\par \noindent 
These are respectively, the universal scalar and gaugino masses,
the universal trilinear and bilinear\footnote{Using renormalisation group
  evolution, the bilinear term is expressed in the low energy MSSM as
  tan$\beta$, the ratio of the vacuum expectation values of the two Higgs
  doublets.} scalar couplings, and the Higgs doublets mass mixing
parameter.

 In this analysis R-parity\footnote{R-parity is a quantum number, defined
  as R = (-1)$^{3(B-L)+2S}$~\cite{PRIMER}, with B, L, and S respectively
  the baryon number, the lepton number and the spin of the
  particle. Non-supersymmetric particles, including the Higgs scalars 
  are R-{\it even}, whilst the supersymmetric particles are all R-{\it
    odd}.} conservation  is also assumed, which leads to three important
phenomenological consequences~\cite{PRIMER}. Firstly, the {\it lightest 
  supersymmetric particle} (LSP) must be absolutely stable. If the 
LSP is electrically neutral, as favoured by cosmological constraints, it
  interacts only weakly with ordinary 
  matter, so escaping detection.
  Secondly, each supersymmetric particle (sparticle) other than
the LSP must eventually decay into a  state which contains an odd number of
LSPs, typically just one. Finally, R-parity conservation implies that
collider experiments could only produce sparticles in even numbers.

Consequently, sleptons could be pair produced at LEP via \ee\ annihilation
into \Zn/$\gamma$ (Figure 1.a). In addition, selectrons can be produced from
t-channel neutralino exchange, which introduces a direct dependence on the
SUSY parameters and the possibility of left and right handed final states
even without mixing via the mass matrix (Figure 1.b).

In a large fraction of the SUSY parameter space the dominant decay of the
\mbox{sleptons ($\tilde{\ell}$)} is to the corresponding lepton flavour plus the
lightest neutralino ($\lsp$) (Figure 2.a), presumed from the MSSM mass 
spectrum to be the LSP. The neutralino  will
escape undetected, hence 
the topology will be characterised by  acoplanar lepton pairs together with
missing  energy.  In most of the analyses described in this paper we search
specifically for such a signature.  

The search can be extended by looking for topologies other than acoplanar
lepton pairs. For certain values of SUSY parameters it is possible for the
second lightest 
neutralino ${\mathrm {\tilde \chi_2^0}}$ to be lighter than the 
  sleptons. If this is the case the slepton can also decay via a cascade to
  a {$\mathrm {\tilde \chi_1^0}$}, with a possible decay chain
  {${\mathrm {\tilde \ell}} \rightarrow \ell\mathrm {\tilde \chi_2^0}$}
  $\rightarrow  \ell\gamma${$\mathrm 
      {\tilde \chi_1^0}$} (Figure 2.b). 
    In order to investigate this channel we have
    searched DELPHI data for events containing acoplanar lepton and photon
    pairs plus missing energy.

\section{Detector description \label{sec:DETEC}}

The {\tt DELPHI} detector and its performance have been described in detail
elsewhere~\cite{DELPHIPER}~\cite{DELPHIPER2}; in the following we present
only a brief description of the components relevant 
to the analyses presented here. 

A system of cylindrical tracking chambers coupled with a 1.2 T uniform
solenoidal magnetic field, directed along the beam axis, enables the
reconstruction of charged particle tracks. The Vertex Detector (VD) consists
of three cylindrical layers of  silicon detectors, at radii  6.3~cm, 9.0~cm
and 11.0~cm. The vertex tracking is aided in the forward regions by
mini-strips and pixel detectors making up the Very Forward Tracker
(VFT)~\cite{VFT} with an angular acceptance between 10$\dgree$ and
25$\dgree$. The Inner 
Detector (ID) is a cylindrical drift chamber (inner radius 12 cm and outer
radius 22 cm). The Time Projection Chamber (TPC), the principal tracking
device  of {\tt DELPHI}, is a cylinder of 30 cm inner radius, \mbox{122 cm} outer
radius and a length of 2.7 m. Each end-plate is divided  into 6 sectors,
with 192 sense wires used for the dE/dx measurement and 16 circular pad rows
used for 3 dimensional  space-point reconstruction. The Outer Detector (OD)
is composed of 24 planks each with 5 layers x 32 columns of drift tubes. The
tubes, 
situated at radii between 196~cm and 207~cm from the beam axis, improve the
precision of the momenta of the charged particles measured by the TPC. In
addition to the barrel tracking, two planes of drift chambers, Forward
Chambers A (FCA) and B (FCB), aligned perpendicular to the beam axis, allowed
tracking in the endcap of the detector, giving a polar coverage
down to 11$\dgree$ and 169$\dgree$ with respect to the $e^-$ beam
direction. 

The electromagnetic calorimetery consists of the High density Projection
Chamber (HPC), covering the barrel region of polar angle $\theta$ in the
range $43\dgree<\theta<137\dgree$, the
Forward ElectroMagnetic Calorimeter (FEMC), consisting of 9064 Cherenkov lead
glass blocks covering $11\dgree<\theta<36\dgree$ 
and  $144\dgree<\theta<169\dgree$, and the STIC (Scintillator TIle
Calorimeter), extending the coverage down to 1.66$\dgree$ from the beam
axis in either direction.  The 40$\dgree$ taggers are a
series of single layer scintillator lead counters used to veto photons and electrons that would otherwise have been missed in the region between
the HPC and FEMC. 

The hadron calorimeter (HCAL) covers 98$\%$ of the solid angle. Muons with
momenta above 2~\GeV\ traverse the HCAL and are recorded in a set of
muon drift chambers; the MUon Barrel (MUB) chambers,  MUon Forward
(MUF) chambers and the Surround Muon Chambers (SMC).

The identification of muons is provided primarily by the algorithm described
in ~\cite{DELPHIPER2}, which relies on the association of charged particles to
signals in the barrel and forward muon chambers. In order to reduce
contamination from cosmic ray particles, the impact parameter with respect to
the beam crossing point was required to be less than 1.5 mm in the $R-\phi$ 
plane. 

Electrons are identified as charged particle tracks with an energy deposit
above 3~\GeV\ in the  electromagnetic calorimeter and with the ratio of the
electromagnetic calorimeter energy to the track momentum from the track above
0.3$c$.  In addition, the shape of the shower profile in the HPC and the 
dE/dx measurement in the TPC were also considered. Forward electrons are 
distinguished from gamma conversions by requiring hits in the VFT. 

A charged particle is identified as a pion if the energy deposited in the
HCAL is greater than 2 GeV, greater than the energy deposited in the 
electromagnetic calorimeters and it does not produce hits in the muon chambers.



\section{Data samples and event generators\label{sec:SAMPLES}}

\vspace{0.1cm}

The total integrated luminosity accumulated by the {\tt DELPHI} experiment
over the two years analysed was 212 pb$^{-1}$. This included 54 pb$^{-1}$ of
data collected at a centre-of-mass energy of 183~\GeV{} and 158
pb$^{-1}$ collected at 189~\GeV.   

Several programs were used to simulate Standard Model (SM) and SUSY
(signal) events in order to estimate background contamination and signal
efficiencies. 

All the models used {\tt JETSET 7.4}~\cite{JETSET} for quark fragmentation
with parameters tuned to represent {\tt DELPHI} data~\cite{DELPHIDAT}.
\vspace{0.1cm} The program {\tt SUSYGEN}~\cite{SUSYGEN} was used to  generate
slepton events and to calculate cross-sections and branching ratios. The
generator {\tt EXCALIBUR}~\cite{EXCALIBUR} was used to model all four-fermion
events,  which includes the coherent interference of all diagrams leading to
a given final state.  For a cross check, {\tt PYTHIA}~\cite{JETSET} was used to
generate samples of  WW, ZZ, We$\nu$ and Zee events.   The processes
\eeto  
$\Zn/\gamma \rightarrow \Pq{}\Paq{}(\gamma)$ were simulated by {\tt PYTHIA}, 
whilst  
the two-fermion backgrounds \eeto $\Zn/\gamma \rightarrow \mu^+\mu^-(\gamma)$
and 
$\tau^+\tau^-(\gamma)$ events were produced by {\tt KORALZ}~\cite{KORALZ}.  The
generators {\tt 
  BABAMC}~\cite{SALVA} and {\tt BHWIDE}~\cite{BHWIDE} were used to simulate
Bhabha scattering.  Two-photon interactions leading to hadronic final states 
were simulated using {\tt TWOGAM}~\cite{TWOGAM} and {\tt BDKRC}~\cite{BDKRC}
for the Quark Parton Model contribution. {\tt BDK}~\cite{BDK} was used for
final states with electrons only, whilst final states with muons or taus were
simulated using {\tt BDKRC}.

Generated signal and background events were passed through a detailed detector
response simulation ({\tt DELSIM})~\cite{DELPHIPER2} and processed with the same
reconstruction and analysis programs as the data. The number of background
events simulated was 
several times larger than the number expected in the data.
\vspace{0.1cm}

\section{Search for selectrons and smuons ($\tilde{\ell}\rightarrow\ell\lsp$)}
    The analysis was performed in two stages. Firstly a loose pre-selection
    was used to obtain a sample containing events with two oppositely charged
    tracks. At this
    stage, various distributions of the real data were compared with
    distributions from simulated SM events.   
 
    After this stage a tighter selection was applied. Tuned
    to both simulated background events and signal events, selections were
    made in order to reduce the expected SM background whilst keeping a
    reasonable efficiency for the signal over a wide range of the   
    $\tilde{\ell}$--$\lsp$ mass combinations.

\subsection{Search for selectrons}

    To search for selectrons, the general topology required was two acoplanar
    electrons and missing energy. The preliminary event selection kept all
    candidates with exactly two well reconstructed oppositely charged
    particles with momentum above 1~\GeVc. One of the two charged particles
    was required to be identified as an electron, rejecting events
    if the other was identified as a muon. At this stage in the analysis, the
    selection consisted mainly of Bhabha and two-photon events, with
    satisfactory 
    agreement observed between data and simulated background
    (Figure~\ref{fig:sele:dmc}).

\vspace {0.1cm}
        To further reduce the SM backgrounds, tighter cuts were
        applied. As two-photon events are predominantly at low polar angles
        and 
        with low momentum it was required that the visible energy be greater 
        than 15~\GeV\ and that the energy deposited in the low angle STIC
        calorimeter be less than 4~\GeV. As a further constraint, the
        invariant mass of the two tracks was required to be greater than 
        4.5~\GeVcc, and the total transverse momentum with respect to the
        beam axis was required to be greater than 5~\GeVc.


   To reduce the number of Bhabha events an upper limit on the visible
   energy of 100~\GeV\ was imposed, whilst also requiring 
   that the neutral energy not associated to the charged tracks be less than
   30~\GeV. Events were also rejected if there were more than four neutral
   clusters in total, each with energy above 0.5~\GeV. Bhabha 
   events are coplanar with a large opening angle,  
   hence it was necessary that the acoplanarity and acolinearity be greater
   than 15$\dgree$.

   Four-fermion events were reduced by the constraints described above, in
   particular the constraint on the visible energy. 

   Contraints were also imposed on the momenta of the two tracks, 
   requiring that both 
   tracks had momenta above 2~\GeVc. It was further required that the
   missing momentum vector pointed to an active region of the detector.
   
    The efficiency for the  signal detection depends on the masses of
    ${\tilde {\mathrm e}}$ and $\XN{1}$. The typical signal efficiency is 
    $\approx$ 50 $\%$.
  
     After this selection a total of 56 candidates were found in the 212
     pb$^{-1}$ of data analysed, compared to 51.2 $\pm$ 1.5 predicted from SM
     processes. Details are given in Table~\ref{tab:sele:bkg}.



\subsection{Search for smuons}

As a pre-selection, exactly two well
reconstructed oppositely charged particles with momenta above 1~\GeVc\ were
required. At least one of the particles had to be identified as a muon. It was
further required that neither track be identified as an electron. The
pre-selection sample consisted mainly of two-photon events, and good
agreement between real data and simulated background was observed
(Figure~\ref{fig:selm:dmc}).

 To further reduce SM backgrounds tighter cuts were applied. It was seen that
 using a sequential cut analysis, the dominant background after a tighter
 selection was W-pair events. 
 These  events become increasingly
 important in 
 regions of high slepton mass and high $\Delta m$
 ($m_{\tilde{\mu}}$-$m_{\lsp}$), where the signal events become virtually
 indistinguishable from 
 the W-pair background. In these regions of  SUSY mass  space the
 cross-section for smuon 
 production is low\footnote{In the selectron scenario the t-channel
 contribution can enhance selectron production for low neutralino masses,
 hence increasing signal sensitivity in the mass regions dominated by W-pair
 backgrounds.}, and hence using  sequential cuts to remove
 this background has a severe effect on the signal sensitivity.    

 Consequently for the 189~GeV data analysis, a different approach was
 adopted to the
 183~\GeV\ analysis, applying a selection procedure which depended on the
 (${\mathrm \tilde{\mu}}$, $\lsp$) mass difference. For a 
 mass difference, $\Delta m$, less than 35~\GeVcc\, where two-photon
 backgrounds are important, an analysis based on sequential cuts was
 performed. For the data taken at 183~\GeV\ this approach was used for the
 full SUSY mass spectrum. 
 However, in the 189~GeV analysis, for regions of $\Delta m$ greater than 
 35~\GeVcc,  where the 
 W-pair backgrounds are kinematically favoured, a probabilistic  analysis
 based on the
 likelihood of an event being compatible with W-pair production was used.  

For regions of $\Delta m$ less than 35~\GeVcc, to remove the two-photon
events, the visible energy was required to be greater than 10~\GeV.  Also
the energy in the STIC had to be less than 1~\GeV. As a further
constraint it was necessary for the invariant mass of the lepton pair to be
greater than 4.5~\GeVcc.  

To remove \eeto \hspace{0.1cm}$\Zn/\gamma \rightarrow \mu ^+\mu^-$ events in
this $\Delta m$ region an upper limit  on the visible energy of 120~\GeV\ was
imposed whilst also 
requiring the unassociated neutral energy to be less than 10~\GeV,  with no
more than two neutral clusters. This background was further suppressed by
accepting only events in which the opening angle between the tracks was less
than $165\dgree$ and the acoplanarity was greater than $15\dgree$.

To reduce W-pair contamination in this low $\Delta m$ region, at a small cost
in signal efficiency, events were rejected if the positively  charged
muon 
was within 40$\dgree$ of the $e^+$ beam direction, or the negatively charged
muon was within 40$\dgree$ of the $e^-$ beam direction.  

For the selection of events kinematically allowed in regions of $\Delta m$
greater than 35~\GeVcc, a discriminating variable was constructed for the
events in the 189~\GeV\ data using
the probability density functions (p.d.f's) of W-pair event variables after the
pre-selection stage. The following variables were chosen due to their high
discriminating power between signal and four-fermion events and their
relatively low correlations:

\begin{itemize}
\item{Product of lepton polar angle and charge (Q cos$\theta$)};
\item{Neutral energy};
\item{Opening angle between the leptons};
\item{Acoplanarity};
\item{Missing energy};
\item{Missing transverse momentum}.
\end{itemize}

\noindent In addition, these variables have excellent agreement between real 
data and simulated background. 

The discriminating function is shown in Figure 5 for data and simulated
background (which is predominantly 2-photon at this stage). 
Also shown is the comparison of the discriminant variable
for a sample of 4-fermion events and a sample of SUSY signal with a $\Delta m$
value close to the W mass (80~\GeVcc). 

Two-photon and di-muon events were removed using the same cuts as in the low
mass window. W-pair contamination was reduced by cutting on the
discriminating function such that signal to background was maximised. It was
further required that the missing momentum vector pointed  towards  active
components of the {\tt DELPHI} detector.   

The efficiency for the signal detection depends on the masses of
${\tilde {\mathrm \mu}}$ and $\lsp$. The cuts used to remove the SM
 background resulted in typical efficiencies of $\approx$ 50 $\%$ for the
regions of low $\Delta m$, and $\approx$ 35 $\%$ for the regions of high
$\Delta m$.

Table~\ref{tab:smu:bkg}  summarises the number of accepted events in the data
together with the predicted number of events from background
sources.  In the data collected at 189~\GeV, for the regions of $\Delta m$$\le$
35~\GeVcc, 17 candidates passed the tight selection, consistent with a
background prediction of 17.5 $\pm$ 0.3 events. For the regions of $\Delta m$
$>$ 35~\GeVcc, 7 candidates remained compared with a background prediction
of 9.2 $\pm$ 0.2 events. In the analysis  of data taken at a centre-of-mass
energy of 183~\GeV,  5 candidates
remained with an expectation of 6.1 $\pm$ 0.6 events from SM processes. 

\vspace{1cm}

\section{Search for staus}
The off-diagonal terms of the slepton mass matrix are proportional to the mass 
of the corresponding SM partners. Important effects caused by these terms 
must be considered when searching for staus. For a certain  mixing between
right and left handed staus, the low-mass stau eigenstate
($\tilde{\tau}_{1}$) can become an important candidate for the lightest
charged supersymmetric particle.  Another consequence of the possible mixing
of staus is the change of the coupling to the \Zn, and consequently
the production cross-section, with the mixing angle.
\subsection{Search for heavy staus}  
The characteristic signature of the production of pairs of heavy staus is
the detection of a $\tau^+\tau^-$ pair with large 
acoplanarity and missing energy. Due to the scalar nature of the stau, the
visible system will tend to be at large angles to the beam. 

Among the SM background processes to this signal are 
s-channel production of tau pairs,
in particular if they arise from a radiative return to the \Zn, with
the ISR photon escaping detection, and 
four-fermion events where the final 
state contains two taus as the only visible particles.
Finally, two photon interactions with $\gamma \gamma \rightarrow
\tau^+\tau^-$ contribute in the case of staus close in mass to the LSP.

To select events with two taus,
well reconstructed charged and neutral particles were first  
collected into clusters of total invariant mass below 5.5~\GeVcc.
Events with exactly two particle clusters (possibly accompanied by 
isolated neutral particles) were considered further if there were no more 
than 6 charged tracks in the event and these gave a total charge of 0 or
$\pm$ 1.  At least two tracks were required to have momentum above 1~\GeVc\:
with one greater than 4~\GeVc. The distributions of data and simulated SM
events agree well at this stage in the analysis, as can be seen in
Figure~\ref{fig:stau:datmc}. 

To ensure that the selected events had the high acoplanarity
typical for the signal, the acoplanarity angle was required to be
above $10\dgree$ ($11\dgree$ for the 183 \GeV\: sample).

Selecting events at high angles to the beam was done by demanding that
at least two charged particles with momentum
above 1~\GeVc\: were observed above 30\dgree\: to the beam axis.
Also, the direction of the vectorial sum of momenta should
be contained in the barrel region at an angle 
greater than $37\dgree$ to the beam
($30\dgree$ in the 183 \GeV\:
sample). 
To reduce the background from radiative returns to the \Zn, none of the 
clusters were allowed to have a total momentum ($p^{\rm JET}$) above 67~\GeVc\:
(60 \GeVc\: in the 183 \GeV\: sample),  the energy of
isolated photons had to be below 20~\GeV\: 
and there should be no signal in any 40\dgree\: tagger. 
Furthermore, the value of the reduced 
centre-of-mass energy  ($\sqrt{s^\prime}$) 
was estimated using the angles of the jets, and 
assuming that a photon was lost in the beam-pipe 
(the triangle rule).
This value
should not fall in the interval  90 to 94 GeV 
(no such cut was made for the 
183 \GeV\: sample). 
In addition, there had to be no  calorimetric energy below $30\dgree$ in
polar angle. This last cut was also very effective against the background
from two-photon events, as it removed all such events where 
either of the initial e$^+$e$^-$ were deflected into the forward
calorimeters.
The two-photon background was also reduced
by discriminating against events where the two clusters were
close together: the acoplanarity angle should not exceed
$170\dgree$ ($176\dgree$ for the 183 \GeV\: sample).

 In order to further suppress the background 
from \eeto \hspace{0.1cm}$\Zn/\gamma \rightarrow \tau ^+\tau^-$ events 
with \mbox{$\tau$-decays} highly asymmetric in visible momentum, the square of the 
transverse momentum with respect to the thrust axis ($\delta$) had to be 
above 0.9~(\GeVc)$^2$. This was the case for the 189~\GeV\: sample; the
condition was more complex in the 183~\GeV\: sample (see below).

At this stage of the analysis, the background
was dominated by the W-pair background, and in order to
further suppress it, the events were analysed 
under the assumption that they were 
indeed W-pair events. The $\theta$ angle 
of the positive W  
($\theta_{W^+}$) was reconstructed by assuming that the direction
of the taus was indentical to that of the jets and applying an unsmearing
procedure (derived from simulated W-pair events) to estimate the true 
momentum of the  taus. The final estimate of $\theta_{W^+}$ was then
given by  the average
of the polar angle of the positive jet, the complement of
the polar angle of the negative jet, and the two 
approximate solutions
to the equation determining the W angle.
As the signal is 
isotropic, whilst the W production is enhanced in the forward-backward
 direction and concentrated at high values of the higher of the two jet
 momenta  ($p^{\rm JET}_{max}$), it was required that the observed values of
$\theta_{W^+}$ (in radians) and $p^{\rm JET}_{max}$ (in~\GeVc\:)
were below the higher of the two lines $\theta_{W^+}$ = 1.5 and
$\theta_{W^+}$ = -0.05 $p^{\rm JET}_{max}$ + 3.7 in
the $p^{\rm JET}_{max} - \theta_{W^+}$ plane.
This cut was applied to the 189~\GeV\: sample;
in the 183~\GeV\: sample, the simpler cut
$\theta_{W^+} \le $ 2.5 radians was used.


This selection was supplemented by cuts that depended on the 
region of the (\msta,\MXN{1}) plane considered, which were tuned to remove the
corresponding backgrounds for that region. 
The sub-division was different in the two samples, as were the
cuts depending on the region.
For the 189~\GeV\: sample, two regions were defined:
$\Delta m$ = \msta-\MXN{1} below or above 20~\GeVcc.
In the region of low mass difference, where the remaining
two-photon background was concentrated,
it was required that the missing transverse momentum ($p_T^{\rm miss}$) 
was greater than 5.4~\GeVc\, whilst, in the 
region of high $\Delta m$, the $p_T^{\rm miss}$ was required to be 
above 8~\GeVc. In the lower (higher) $\Delta m$ region, it was also required 
that the  highest momentum of any identified lepton in the event was less 
than 20~\GeVc\: (22 \GeVc),
in order to further suppress the remaining W-pair background.

For the 183~\GeV\ sample three regions of $\Delta m$ were considered:
less than 22~\GeVcc\, \mbox{$22-50$~\GeVcc\ } and more than 50~\GeVcc.
In these three regions,  $\delta$ was required to exceed 0.4,
1.0 and 0.4 (GeV/$c$)$^2$ respectively, and $p_T^{\rm miss}$ to exceed
5.5, 6, and 6~\GeVc, respectively.  
The momentum of any identified lepton in the event should be less than
30~\GeVc\: (independent of $\Delta m$).

Table~\ref{tab:stau:bkg} summarises the number of accepted events in the 
data for the different selections together with the expected numbers of 
events from the different background channels. In the 212 pb$^{-1}$ data 
sample analysed, 16 candidates were found, with a background estimation 
of 18.1 $\pm$ 0.8\: from SM processes. The signal detection efficiency was 
of the order of 20$\%$ for the 189~\GeV\ sample and 30$\%$ for the 183~\GeV\
sample.   
 
\subsection{Search for light staus without coupling to the Z$^o$}  
To a large extent a light stau can be excluded using the agreement 
of the decay width of the \Zn\ resonance with the SM prediction, 
as observed at LEP1~\cite{ZWIDTH}.  The corresponding cross-section 
limit of 150~pb for non-standard processes at $\sqrt{s} = M_{\mathrm Z}$ 
excludes a $\sta_R$ below 25~\GeVcc. However, at the stau mixing angle 
giving the minimum cross-section, the coupling to the \Zn\
vanishes and no exclusion is possible using this method. The high mass
analysis described in the previous section loses its efficiency for stau
masses below 20~\GeVcc. This is mainly due to the
fact that the stau-pairs are highly boosted at such a low mass, so
that they fail the acoplanarity cut.
Therefore a specific search was required for \msta\: in the range 
from $m_{\tau}$ to 27~\GeVcc\: and this mixing angle.    
After selecting two tau events as described in the previous section,
it was required that there was no identified isolated photon in the event.
To reduce the background from Bhabha
scattering, we demanded that the acolinearity angle was  
above 0.4\dgree\ and
$p^{\rm JET}_{max}$ below 70~\GeVc.   
Furthermore, the missing transverse momentum in the event was 
required to be above 6~\GeVc, the angle of the most energetic track 
in each hemisphere of the detector to be
above 50\dgree\: to the beam,
and the direction of the vectorial sum of momenta should
be above a polar angle of 40\dgree.
To further reduce 
the background from radiative return to the \Zn, 
the value of 
$\sqrt{s^\prime}$ should not be between 
82 to 102~\GeV\: \footnote{It should be noted that as a consequence of the 
vanishing coupling to the \Zn, radiative return does not occur
in  \sta\: production at the mixing angle yielding the minimal cross-section,
even at \msta\: below $M_Z$/2.}.

For \msta\: above 15~\GeVcc, the cut 
on acoplanarity was more restrictive:
it was demanded to be above 
$4 \dgree$.

With these cuts, a total of 122 events were selected in the mass region
below 15~\GeVcc\: and 50 events in the higher region.
The SM background was estimated to 150.1 $\pm$ 2.3 and 55.7 $\pm$ 1.7 in the
two regions (Table~\ref{tab:stau:bkg2}).
The efficiencies were practically flat for both
mass ranges.  In the first case, the efficiency was around
25\%, decreasing to 10\% for \mbox{$\Delta m$ $=m_\tau$}. 
In the second case it was
around 15\%, decreasing to $\le$ 2 \% for \mbox{$\Delta m$$ < 2\:\GeVcc$}.

\section{Search for cascade decays}
In order to extend the slepton search, topologies from
cascade decays of the $\tilde{\ell}$ have been considered. There are regions
of the SUSY parameter  space  where the sleptons may also 
decay into the $\neutrala$ plus the corresponding lepton
($\tilde{\ell}\rightarrow\ell\nlsp$).  

For smuons it presents some
advantages as the dependence of the smuon mass limit on the SUSY parameter
$\mu$ is considerably reduced. The 
$\neutrala$ may decay to $\neutralino\gamma$, with  the $\neutralino$
being the LSP which escapes undetected.  The topology for these events is 
acoplanar lepton pairs and  two photons plus missing energy.    The main
advantage  of searching for this type of event is that the experimental
signature is very clean with very small SM backgrounds since the
emission of  photons requires higher orders and  extra $\alpha_{EM}$ factors. 
However  the  cascade  decay may be suppressed as the branching ratio
$BR(\neutrala\rightarrow \neutralino\gamma)$ may be small. The
$BR(\neutrala\rightarrow 
\neutralino\gamma)$ is close  to 1 when both neutralinos have a
similar mass, but in this near degenerate case the outgoing photon 
has low energy, so making detection difficult.

The search for these events was done in a two step procedure. First, samples
 of  $ee\gamma\gamma$ and
$\mu\mu\gamma\gamma$ events were selected following  loose cuts. Before
applying a tighter selection a  likelihood function was defined  for the main
background channels (one for tagging Bhabha events  in the $ee\gamma\gamma$
sample and the second one for tagging the $e^+e^- \rightarrow \mu^+\mu^-$
entering the $\mu\mu\gamma\gamma$ samples) contributing to the sample. Then
the tight selection is combined with  the likelihood in order to remove those
events compatible with the SM processes. 

\subsection{\boldmath $ee\gamma\gamma$ selection}
A sample of $ee\gamma\gamma$ events  was selected requiring
 only two charged  tracks reconstructed  with momentum  greater
than  3~\GeVc{} and at least two photons reconstructed with energy above
 1~\GeV. If several photons were selected, only the most energetic two were
 retained. Events were rejected if either of the charged tracks was
 consistent with positive pion or muon identification (see section 2).   An
 acceptance cut  demanding 
 that all 4 particles should lie in the $10^\circ<\theta<170^\circ$ region was 
applied in order to remove most of the  two-photon interactions.  The same
cut was applied to the missing momentum vector, since this can be close to
the beam axis for events coming from a radiative return to the $\Zn$. Finally
 the  acolinearity of the  electron and photon pairs  had to be larger 
than 3\dgree. After this stage in the analysis, reasonable agreement was
observed between real data and simulated SM processes contributing to the
sample (Figure~\ref{fig:eegg_1}).

The Bhabha likelihood was  built  according to  the probability density
functions (p.d.f's) of the  visible
energy, invariant  mass between the electron-photon pair with smallest opening
angle,  the invariant mass of the electron pair  and most energetic photon,
and the angle between the missing   momentum  and     the  closest
electron   and photon.   

A sample with most SM events removed was  selected by demanding that  the
events satisfying the loose   $ee\gamma\gamma$ selection comply  with  the
following  cuts: acolinearity and acoplanarity  of   the electron and  photon
pairs  above  $6^\circ$ and a Bhabha  probability less than
3\%.  According to this selection, $4.2\pm0.9$ events were expected from
the SM and 5 seen (\mbox{Table \ref{tab:eegg_2}}).

The efficiency for signal detection depended on the lepton and photon energies.
An efficiency map was computed for a range of points of the SUSY parameter
space. Typical efficiencies for detection were of the order of 40$\%$.

\subsection{\boldmath $\mu\mu\gamma\gamma$ selection}
The selection of a sample of $\mu\mu\gamma\gamma$ events proceeded in 
a similar manner to that of the $ee\gamma\gamma$ events. It was required that
at least one of the charged tracks had to be identified as a muon, and neither
of them as an electron. The rest of the kinematic cuts were the same as
mentioned 
in the $ee\gamma\gamma$ selection. Events with more than two photons were
further considered if the energy of the extra photon(s) was below 10~\GeV.

The $e^+e^- \rightarrow  \mu^+\mu^-$ likelihood was built from the p.d.f.s of
the visible energy, momentum of the leading muon, invariant mass of the muon
pair with the most energetic photon, the angle between the missing momentum 
and the closest muon, and the opening angle of the muon pair.

The  tight  $\mu\mu\gamma\gamma$ selection consisted basically  of the same
topological cuts as the tight $ee\gamma\gamma$ selection. In addition to
these cuts, it was required that the likelihood for an event being
consistent with $e^+e^- \rightarrow \mu^+\mu^-$ be less than  5\%.  With these
cuts 2.9 $\pm$ 0.7  events were expected from the SM and 
 3 seen in the data (Table \ref{tab:uugg_2}). The cuts used to remove SM
background resulted in typical efficiencies for signal detection of 45$\%$.



\section{Results \label{sec:RESULT}}

Limits on slepton masses can be derived using several
different assumptions. Scalar mass unification suggests
lower masses and cross-sections for the partners of
right handed fermions. Hence we have assumed that only right handed 
selectrons ($\selr$) and smuons ($\smur$)  are produced, leading to
conservative mass limits. 

For third generation sfermions, Yukawa couplings can be large,
leading to an appreciable mixing between the pure weak hypercharge
states. The production cross-section depends on this mixing, due
to the variation in strength of the coupling to the $\Zn$ component of the weak
current, and has a minimum at a mixing angle of $42\dgree$.
Consequently, the results for the stau analyses are presented under these two
assumptions; right handed stau production ($\staur$), and minimal mixing
stau production ($\staum$). 

\vspace{1cm}
\section{Exclusion limits}
Exclusion limits for slepton pair ($\tilde{\ell}\tilde{\ell}$) production were
obtained, taking into account the signal efficiencies for each $\tilde
{\ell}-\lsp$ mass point, the cross-section and branching ratios for slepton
production, and the number of data and background events kinematically
compatible for a given mass combination. Signal events have been generated
assuming model input values of \mbox{$\tan\beta$ = 1.5} and 
\mbox{$\mu$ = -200 \GeVcc{}}. The limits were calculated using a
likelihood ratio method described in~\cite{ALRMC}. Expected exclusion zones
were calculated using the same algorithm, from simulated background-only
experiments. 

Figure~\ref{fig:all:xcl}.a shows the 95\% CL exclusion regions for
$\selr\selr$ production, obtained using the full 212 pb$^{-1}$ of data. For
the selectrons, we exclude masses up to $m_{\sel_R} \le$ 87~\GeVcc, providing the
mass difference between the selectron and the LSP is above 20~\GeVcc.
 
Figure~\ref{fig:all:xcl}.b shows the 95\% CL exclusion regions for
$\smur\smur$ production, obtained by combining the 183~\GeV{} data with the 
189~\GeV{} data. For the  smuons, we exclude masses up to $m_{\smu_R} \le$ 
80~\GeVcc, providing the mass difference between the smuon and the LSP is
above 5~\GeVcc.

Exclusion limits on $\sta\sta$ production were obtained taking into account
the signal efficiencies for each ${\sta} - \lsp$ mass point.
When determining whether data or background events were kinematically
compatible with the mass point, the end point of the expected momentum
spectrum of the visible reconstructed tau was
used. Figure~\ref{fig:all:xcl}.c shows the 95\% CL 
$\sta_R$ exclusion  region  obtained by combining the previous data at lower
energies with the 183~\GeV{} and 189~\GeV{} data,
and Figure~\ref{fig:all:xcl}.d shows the exclusion regions in the case of 
the mixing angle yielding the minimal cross-section. For the staus, a mass
limit can be set at 73 to 75~\GeVcc{} (depending 
on mixing) for mass differences between the stau and the LSP above 
\mbox{10~\GeVcc}.
The dedicated search for a low-mass stau yields that a $\sta$
of a mass below 12.5~\GeVcc{} is excluded at 95\% CL for any mixing angle,
provided that $\Delta m$ is greater than $m_{\tau}$.

For the cascade decay analysis, assuming $\tilde{e}_R\tilde{e}_R$
or $\tilde{\mu}_R\tilde{\mu}_R$ production, one can set exclusion regions in
the SUSY parameter  space. To set limits, one has to consider the cross
section for the selectron or smuon production and the branching ratios
for the  ${\tilde{\ell} \rightarrow \ell  \neutrala}$ and
 $\neutrala \rightarrow  \neutralino
\gamma$ must be taken into account.  These cross-sections and  branching
ratios depend on the actual 
values of the SUSY parameters. The excluded regions for a given value of the
common scalar mass, ${\tilde{m}}_0$, and  $\tan\beta$ are presented in
Figure(~\ref{fig:exclusion2}) as a function of the Higgs superfield mass
parameter $\mu$ and ${\tilde{m}}_{1/2}$.   


\section{Conclusions \label{sec:CONCLU}}

In a data sample of 212 pb$^{-1}$ collected by the {\tt DELPHI} detector
at centre-of-mass energies of 183~\GeV\ and 189~\GeV, searches were performed 
for events with acoplanar lepton pairs. The mass limits produced assume
input parameters for the Higgs mass mixing parameter, $\mu$, of -200~\GeVcc\
and ratio of the vacuum expectation value of the Higgs doublets, tan$\beta$,
of 1.5.

For the selectron pairs, 56 candidates remained after selection, with an
expectation of 51.2 $\pm$ 1.5 from SM processes. This allowed a lower limit
on the mass for the $\sel_R$ to be set at 87~\GeVcc\ for $\Delta m$ $>$ 
20~\GeVcc.

In the search for smuon production at 189~\GeV, 17 events were selected
for regions of low $\Delta m$, with 17.5 $\pm$ 0.3 expected
from Standard Model processes. For regions of $\Delta m$ $>$ \mbox{35~\GeVcc}, 7
candidates were selected, with a background expectation of \mbox{9.2 $\pm$ 0.2}
events. At 183~\GeV, 5 candidates were selected with a background expectation
of \mbox{6.1 $\pm$ 0.6} events. Combining these data, a mass limit for $\smu_R$ of 
80~\GeVcc\: was obtained for  \mbox{$\Delta m$ $>$ 5 \GeVcc\:}.    

In the search for stau production, 7 events were selected at 183~\GeV{} with 
7.5 $\pm$ 0.5 expected from SM processes. At 189~\GeV, 9 candidates passed
the selection criteria with a background of 10.6 $\pm$ 0.7
expected. Combining this data with all our 
previous data at lower energies\cite{183PUB}, a
mass limit  for the stau can be set at  75~\GeVcc\ if the stau is purely
a partner to the right handed tau, and at 73~\GeVcc\ if the
stau mixing angle is such that the production cross-section is minimal.  

In the search for a low-mass stau, 122 events were selected in the mass region
below 15~\GeVcc\: and 50 events in the region between 15~\GeVcc\: and 27~\GeVcc\:.
The background was 150.1 $\pm$ 2.3 and 55.7 $\pm$ 1.7 in the
two regions.
Combining these results with all previous data ~\cite{183PUB},
a $\sta$ with mass below 12.5~\GeVcc\ can be excluded for any mixing angle,
provided that $\Delta m$ is greater than $m_{\tau}$.

Events with the topology of $\ee\gamma\gamma$, $\mu^+\mu^-\gamma\gamma$, and
missing energy were analysed and a search for ${\mathrm
  {\tilde \ell}} \rightarrow \ell{\mathrm {\tilde \chi_2^0}} \rightarrow
\ell\gamma{\mathrm   {\tilde \chi_1^0}}$ was performed. For the cascade decay
selectron search 5 events were selected with an expectation of 4.2 $\pm$ 0.9
from SM process. For the smuon case 3 events remained with an expectation of
2.9 $\pm$ 0.7. No excess over the Standard Model prediction was found. 

\subsection*{Acknowledgements}
\vskip 3 mm
 We are greatly indebted to our technical 
collaborators, to the members of the CERN-SL Division for the excellent 
performance of the LEP collider, and to the funding agencies for their
support in building and operating the DELPHI detector.\\
We acknowledge in particular the support of \\
Austrian Federal Ministry of Science and Traffics, GZ 616.364/2-III/2a/98, \\
FNRS--FWO, Belgium,  \\
FINEP, CNPq, CAPES, FUJB and FAPERJ, Brazil, \\
Czech Ministry of Industry and Trade, GA CR 202/96/0450 and GA AVCR A1010521,\\
Danish Natural Research Council, \\
Commission of the European Communities (DG XII), \\
Direction des Sciences de la Mati$\grave{\mbox{\rm e}}$re, CEA, France, \\
Bundesministerium f$\ddot{\mbox{\rm u}}$r Bildung, Wissenschaft, Forschung 
und Technologie, Germany,\\
General Secretariat for Research and Technology, Greece, \\
National Science Foundation (NWO) and Foundation for Research on Matter (FOM),
The Netherlands, \\
Norwegian Research Council,  \\
State Committee for Scientific Research, Poland, 2P03B06015, 2P03B1116 and
SPUB/P03/178/98, \\
JNICT--Junta Nacional de Investiga\c{c}\~{a}o Cient\'{\i}fica 
e Tecnol$\acute{\mbox{\rm o}}$gica, Portugal, \\
Vedecka grantova agentura MS SR, Slovakia, Nr. 95/5195/134, \\
Ministry of Science and Technology of the Republic of Slovenia, \\
CICYT, Spain, AEN96--1661 and AEN96-1681,  \\
The Swedish Natural Science Research Council,      \\
Particle Physics and Astronomy Research Council, UK, \\
Department of Energy, USA, DE--FG02--94ER40817. \\

\clearpage
\newpage


\newpage
\clearpage
\newpage
\newpage
{\setlength{\extrarowheight}{3pt}
\begin{table}[hbt]
  \begin{center}
    \begin{tabular}{|*{3}{c|}}
      \hline
      $\sqrt{s}$ (GeV)         & 183              & 189 \\ \hline
      Observed events     & 11               &  45 \\ \hline 
      Total background    & 12.7 $ \pm$ 0.8  & 38.5 $\pm$ 1.3  \\ 
      \hline \hline 

      $\Zn/\gamma \to (\mu \mu, $ee$, \tau 
      \tau) (n\gamma)$    & 1.7 $ \pm$   0.2    &  1.4 $\pm$ 0.1  \\ \hline
      4-fermion events    & 10.5 $ \pm$  0.8   & 34.2 $\pm$ 1.3  \\ \hline
      $\gamma\gamma \to $ee$ , \mu\mu, \tau\tau$ 
                          & 0.5 $\pm$ 0.1    & 2.9  $\pm$ 0.1 \\ \hline
    \end{tabular}
  \end{center}
  \caption[.]{
    Selectron candidates, 
    together with the total number of background events expected
    and the contributions from major background sources. Results shown are
    for 54 pb$^{-1}$ of data analysed at 183~\GeV and \mbox{158 pb$^{-1}$} of
    data 
    analysed at 189~\GeV. 
    \label{tab:sele:bkg}
    }
\end{table}
}



{\setlength{\extrarowheight}{3pt}
\begin{table}[hbt]
  \begin{center}
    \begin{tabular}{|*{4}{c|}}
       \hline
       & \multicolumn{2}{c|} {189 GeV} & {183 GeV} \\ \hhline{|~|-|-|-|}
       & $\Delta m$ $\le$ 35 GeV/$c^2$ & $\Delta m$ $>$ 35 GeV/$c^2$ & All regions\\ \hline
       Observed events & $17$ & 7 & $5$\\ \hline 
       Total background & 17.5 $\pm$ 0.3  & 9.2 $\pm$ 0.2 & 6.1 $\pm$ 0.6\\
       \hline \hline  
       $\Zn/\gamma \to (\mu \mu, $ee$, \tau 
       \tau) (n\gamma)$ & 0.9 $\pm$ 0.1 & 1.6 $\pm$ 0.2 & 0.2 $\pm$ 0.1\\
       \hline 
       4-fermion events & 15.7 $\pm$ 0.3 & 5.5 $\pm$ 0.1 & 5.8 $\pm$ 0.6\\
       \hline 
       $\gamma\gamma \to $ee$ , \mu\mu, \tau\tau$ & 0.9 $\pm$ 0.1  & 2.1 $\pm$
       0.2 & 0.1 $\pm$ 0.1 \\ \hline
    \end{tabular}
  \end{center}
  \caption[.]{
    Smuon candidates, together with the total number of background events
    expected and the contributions from major background sources. The results
    are shown for the two regions of the SUSY mass space analysed 
    at a centre-of-mass energy
    of 189~\GeV, and the full mass spectrum analysed at a centre-of-mass
    energy of 183~\GeV. 
    \label{tab:smu:bkg}
    }
\end{table}
}

{\setlength{\extrarowheight}{3pt}

\begin{table}[hbt]

\begin{center}

\begin{tabular}{|*{3}{c|}}

\hline

 $\sqrt{s}$ (GeV)         & 183              & 189 \\ \hline

 Observed events          &   7              & $9$ \\ \hline 

Total background          & 7.5 $\pm$ 0.5      & 10.6 $\pm$ 0.7 \\ \hline \hline 

$\Zn/\gamma \to (\mu \mu, $ee$, \tau 

\tau, $\Pq{}\Paq{}$) (n\gamma)$ 

                          &  1.0  $\pm$ 0.3    & 1.5 $\pm$ 0.2 \\ \hline

4-fermion events          &  5.0  $\pm$ 0.3    & 7.2 $\pm$ 0.4 \\ \hline

$\gamma\gamma \to \tau^+\tau^-$  

                          &  0.9 $\pm$ 0.2     & 0.8 $\pm$ 0.3 \\ \hline

$\gamma\gamma \to $ee$, \mu\mu$, $\Pq{}\Paq{}$ 

                          & 0.6  $\pm$ 0.1     & 1.1 $\pm$ 0.5 \\ \hline

\end{tabular}

\end{center}

\caption[.]{

Stau candidates in the search for high mass staus, together with the total number of background events expected

and the contributions from major background sources

for centre-of-mass energies of 183~\GeV and 189~\GeV.\label{tab:stau:bkg}

}

\end{table}

\begin{table}[hbt]

\begin{center}

\begin{tabular}{|*{5}{c|}}

\hline

                          & \multicolumn{2}{c|}{$\msta < 15\ \GeVcc$}
                          & \multicolumn{2}{c|}{$15\ \GeVcc \leq \msta < 27\ \GeVcc$} \\

\hline

 $\sqrt{s}$ (GeV)         & 183              & 189  & 183              & 189 \\ 

\hline

 Observed events          &  31              &  91  &  12              & 38 \\ 

\hline 

Total background          & 38.8 $\pm$ 1.5   & 111.3  $\pm$ 1.8 

                                 &14.4 $\pm$ 0.9   & 41.3 $\pm$ 1.4 \\ 

\hline \hline 

$\Zn/\gamma \to (\mu \mu, 

$ee$, \tau  \tau, 

$\Pq{}\Paq{}$) (n\gamma)$ 

                          &  30.9  $\pm$ 1.4   &  86.5 $\pm$ 1.5  

                                  &  7.6  $\pm$ 0.7   & 19.9 $\pm$ 0.7  \\ 

\hline

Bhabha                    &  1.1  $\pm$ 0.2  & 3.3  $\pm$ 0.7  

                                  &  0.3 $\pm$ 0.1    &  0.9 $\pm$ 0.4 \\ 

\hline

4-fermion events          &   3.9  $\pm$ 0.3  & 12.8 $\pm$ 0.5  

                                  &  3.9  $\pm$ 0.3  & 12.6 $\pm$ 0.4  \\ 

\hline

$\gamma\gamma \to \tau^+\tau^-$,

  ee $, \mu\mu$, \Pq{}\Paq{} 

                          & 2.9  $\pm$ 0.4   &  8.7 $\pm$ 0.6  

                                  & 2.6  $\pm$ 0.4   & 7.9  $\pm$ 1.1  \\ 

\hline

\end{tabular}

\end{center}

\caption[.]{

Stau candidates in the search for low-mass staus, 

together with the total number of background events expected

and the contributions from major background sources

for centre-of-mass energies of 183~\GeV{} and 189~\GeV.\label{tab:stau:bkg2}

}

\end{table}

}

{\setlength{\extrarowheight}{3pt}

\begin{table}[t]
  \begin{center}
    \begin{tabular}{|*{2}{c|}}
      \hline
      Observed events & $5$ \\       
      \hline 
      Total background & $4.2 \pm 0.9$ \\ 
      \hline \hline 
      Bhabha  &  2.0 $\pm$ 0.4  \\      \hline 
      $\gamma\gamma\rightarrow \tau^+\tau^-$& 1.3 $\pm$ 0.7  \\      \hline 
      $e^+e^-\rightarrow \tau^+\tau^-$& 0.6 $\pm$ 0.3  \\      \hline 
      4-fermion& 0.3 $\pm$ 0.3  \\      \hline 
    \end{tabular}
  \end{center}
  \caption[.]{Break down of the individual contributions to the tight
  $ee\gamma\gamma$ sample. Results presented are for data taken at
  centre-of-mass energies of 183~\GeV{} and 189~\GeV.
    \label{tab:eegg_2}
    }
\end{table}
}

{\setlength{\extrarowheight}{3pt}
\begin{table}[hbt]
  \begin{center}
    \begin{tabular}{|*{2}{c|}}
      \hline
      Observed events & $3$ \\ 
      \hline 
      Total background & $2.9 \pm 0.7$ \\ 
      \hline \hline 
      $e^+e^-\rightarrow \mu\mu$  &   0.9 $\pm$ 0.4  \\      \hline 
      $e^+e^-\rightarrow \tau\tau$& 1.2 $\pm$ 0.5  \\      \hline 
      4-fermion events & 0.8 $\pm$ 0.3  \\     
      \hline
    \end{tabular}
  \end{center}
  \caption[.]{Break down of the individual contributions to the tight
  $\mu\mu\gamma\gamma$ sample. Results presented are for data taken at
  centre-of-mass energies of 183~\GeV{} and \mbox{189~\GeV}.
    \label{tab:uugg_2}
    }
\end{table}
}

\clearpage
\newpage
\begin{fmffile}{selpics}
  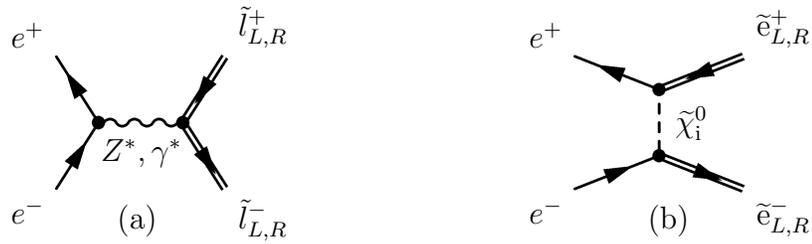
\begin{figure}[!htbp]
    \begin{center}
      \mbox{\fmfframe(0,5)(0,15){\begin{fmfgraph*}(80,50)
            \fmflabel{$e^{-}$}{em} \fmflabel{$e^{+}$}{ep} \fmfleft{em,ep}
            \fmf{fermion,tension=2}{em,Zee,ep}
            \fmflabel{$\tilde{l}^{-}_{L,R}$}{s} 
            \fmflabel{$\tilde{l}^{+}_{L,R}$}{sb}
            \fmf{dbl_plain}{sb,Zss,s}\fmf{phantom_arrow}{sb,Zss,s}
            \fmfright{s,sb} \fmf{photon,tension=2,label=$Z^*,,\gamma^*$}{Zee,Zss}
            \fmfdot{Zee,Zss}
          \end{fmfgraph*}}}\hspace{3cm}
      {\hspace{-5cm} (a) \hspace{5cm}}
      \mbox{\fmfframe(0,5)(0,15){\begin{fmfgraph*}(80,50)
            \fmflabel{$e^{-}$}{em} \fmflabel{$e^{+}$}{ep}
            \fmflabel{$\PSe^{-}_{L,R}$}{s} \fmflabel{$\PSe^{+}_{L,R}$}{sb}
            \fmf{fermion,tension=2}{epxsb,ep} \fmf{dbl_plain}{sb,epxsb}
            \fmf{phantom_arrow}{sb,epxsb} \fmf{fermion,tension=2}{em,emxs}
            \fmf{dbl_plain}{emxs,s}\fmf{phantom_arrow}{emxs,s}
            \fmf{dashes,tension=2,label=$\PSgxz$}{emxs,epxsb} \fmfleft{em,ep}
            \fmfright{s,sb} \fmfdot{epxsb,emxs}
          \end{fmfgraph*}}}{\hspace{-1.7cm}\vspace{.8cm} (b)
      \hspace{2cm}\vspace{-0.8cm}} 
      \caption{Production diagrams for sleptons in the MSSM. (a) Shows the
        pair-production of sleptons, a possible scenario at collider
        experiments. (b) Shows the additional t-channel contribution to
        selectron production.}
    \end{center}
  \end{figure}
\end{fmffile}

\begin{fmffile}{decay}
  \begin{figure}[!htbp]
    \begin{center}
      \mbox{\fmfframe(0,5)(0,15){\begin{fmfgraph*}(80,50)
            \fmflabel{$\tilde{\mathrm \ell}$}{s}
            \fmfleft{s}
            \fmf{dbl_plain}{s,d}\fmf{phantom_arrow}{s,d}\fmf{fermion}{d,f1}
            \fmf{dashes}{d,f2}\fmflabel{$\ell$}{f1}\fmflabel{$\lsp$}{f2}
            \fmfright{f2,f1} \fmfdot{d}
          \end{fmfgraph*}}}\hspace{3cm}
      {\hspace{-5cm} (a) \hspace{5cm}}
      \mbox{\fmfframe(0,5)(0,15){\begin{fmfgraph*}(80,50)
            \fmfleftn{i}{1}\fmfrightn{o}{3}\fmfdot{v1}\fmfdot{v2}
            \fmflabel{$\tilde {\mathrm \ell}$}{i1}
            \fmflabel{$\ell$}{o3}
            \fmflabel{$\gamma$}{o2}
            \fmflabel{$\lsp$}{o1}
            \fmf{dbl_plain_arrow}{i1,v1}
            \fmf{fermion}{v1,o3}
            \fmf{dashes,label=$\neutrala$}{v1,v2}
            \fmf{dashes}{v2,o1}
            \fmf{boson,tension=.5}{v2,o2}
          \end{fmfgraph*}}}{\hspace{-1.7cm}\vspace{.8cm} (b)
      \hspace{2cm}\vspace{-0.8cm}} 
      \caption{Slepton decay diagrams. (a) Shows the slepton decaying into
        a lepton of same flavour and the LSP. (b) Shows the cascade decay;
        the slepton decaying into the lepton plus the second lightest
        neutralino, followed by a radiative decay to the LSP.}
    \end{center}
  \end{figure}
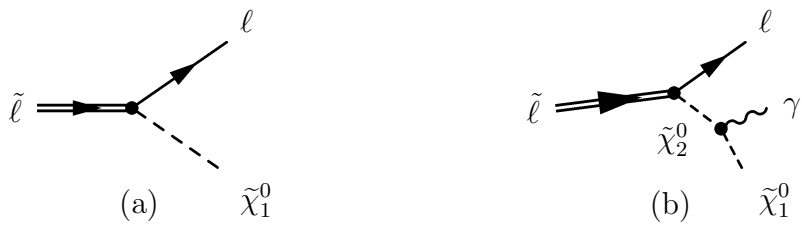
\end{fmffile}

\vspace{0.2cm}
\begin{figure}[htbp]
\centerline{\epsfxsize=16.0cm \epsfysize=18.0cm \epsfbox{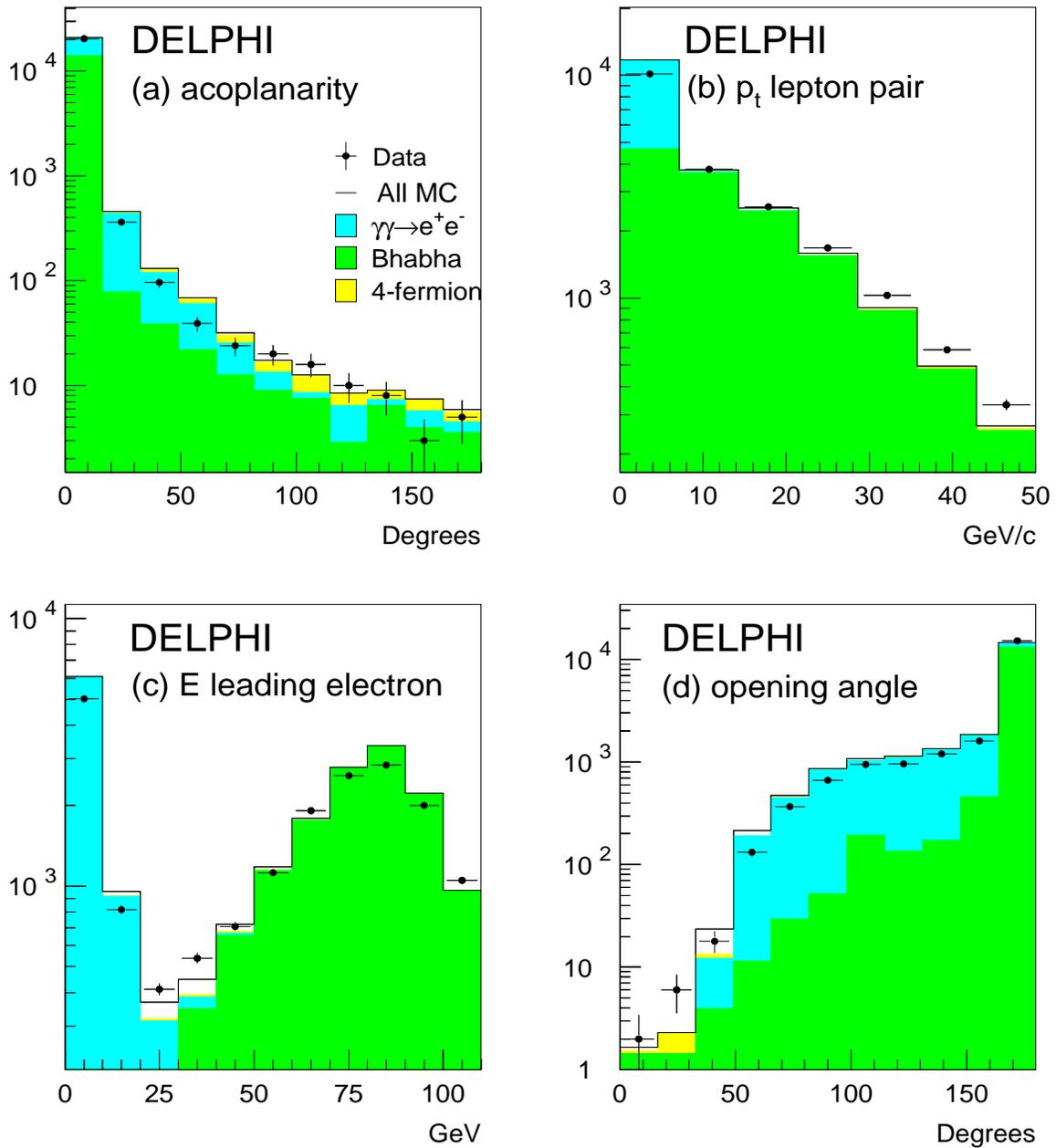}}
\caption{A pre-selection comparison of data and simulated SM events in
  the selectron 
  analysis at 189~\GeV. The plots show; (a)
  Electron pair acoplanarity, (b) Transverse momentum of the electron pair,
  (c) Energy of leading electron, (d) Opening angle between lepton pair. The
  dots with   error bars show the data, while the simulation is plotted as a
  histogram.}  
\label{fig:sele:dmc}
\end{figure}

\begin{figure}[htbp]
\centerline{\epsfxsize=16.0cm \epsfysize=18.0cm \epsfbox{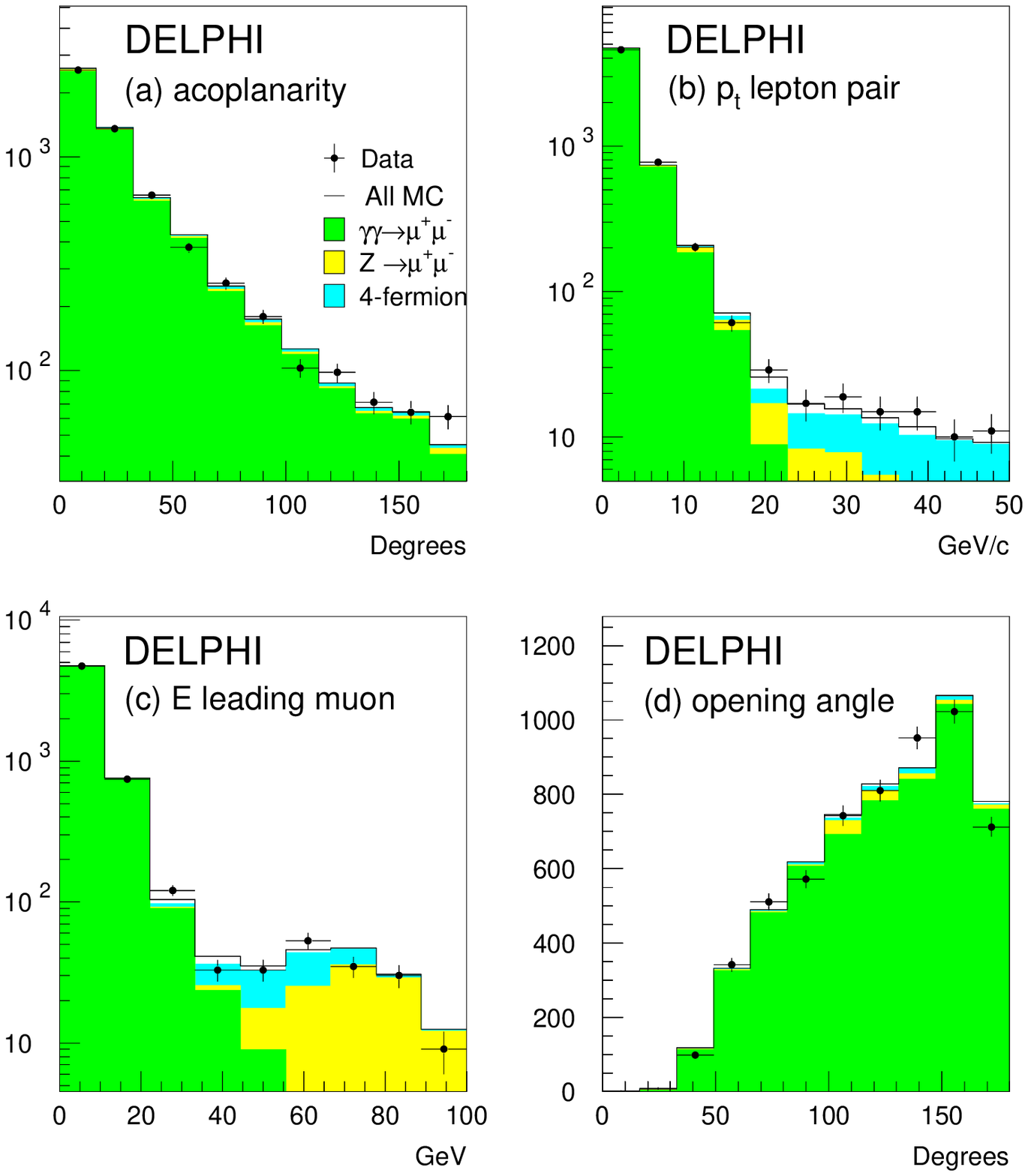}}
\caption{A pre-selection comparison of data and simulated SM events in
  the smuon analysis at \mbox{189~\GeV}. The plots show;  (a)
  Muon pair acoplanarity (b) Transverse momentum of the muon pair, (c) Energy
  of leading muon, (d) Opening angle of the muon pair. The dots with error
  bars show the data, while the simulation is plotted as a histogram.}   
\label{fig:selm:dmc}
\end{figure}

\begin{figure}[htbp]
\mbox{\epsfxsize=8.0cm \epsfysize=8.0cm\epsfbox{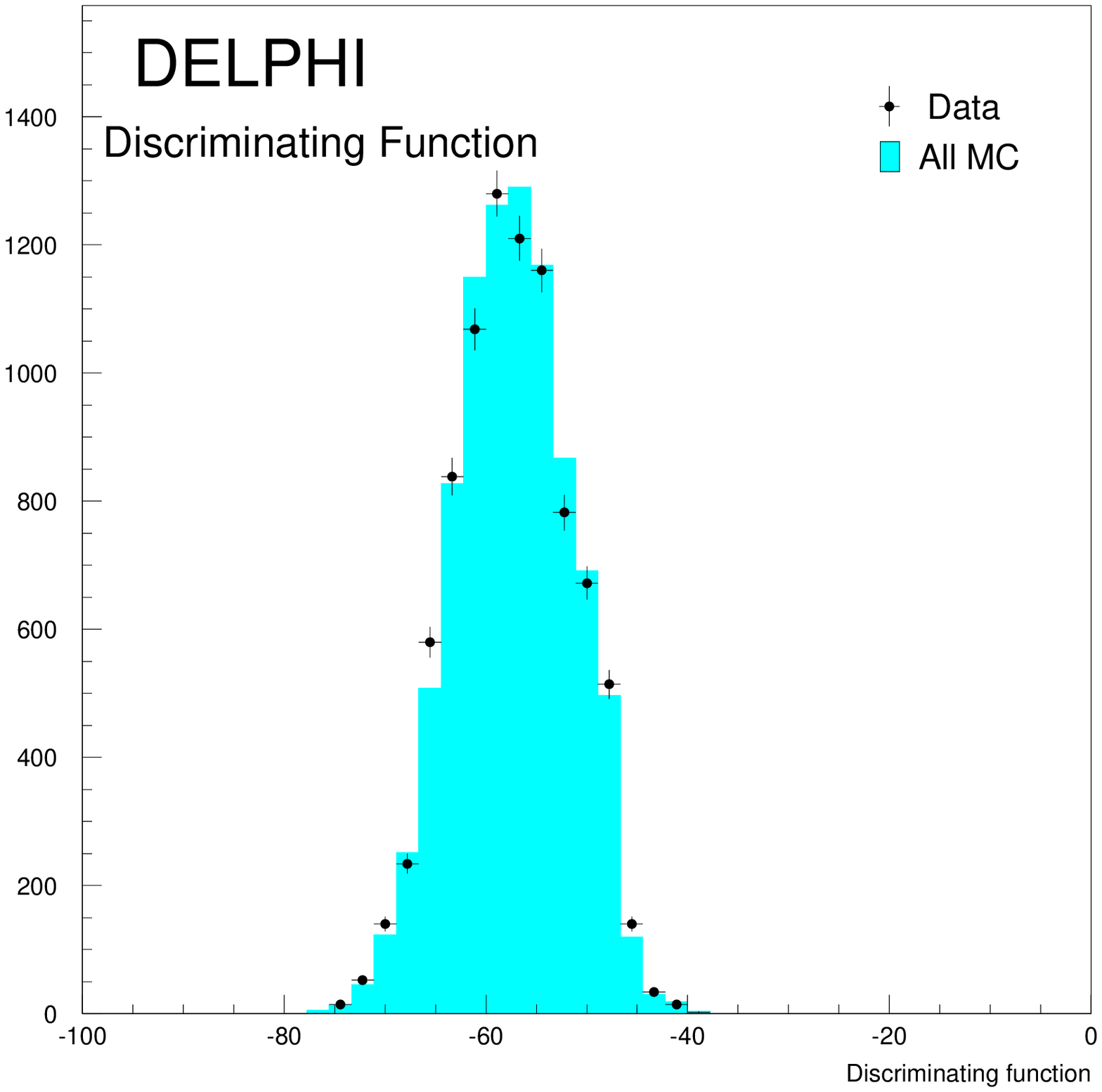} 
\epsfxsize=8.0cm \epsfysize=8.0cm\epsfbox{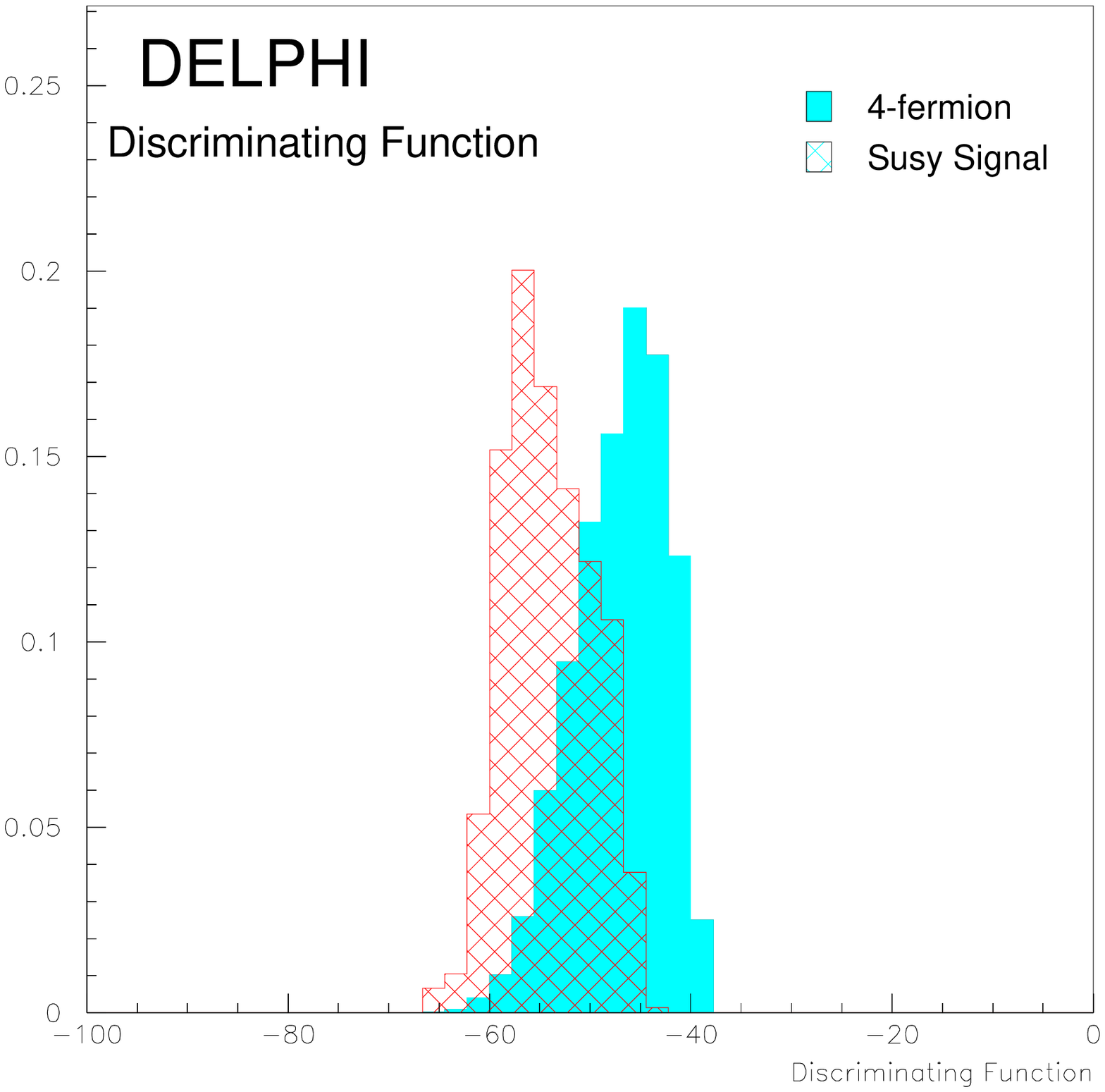}}
\label{fig:dishat}
\caption{Function used to discriminate against W-pair
  backgrounds. Left: The points represent the data and the solid histogram
  shows the contribution from Standard Model backgrounds. Right: The
  discriminating function for a sample of 4-fermion events (solid) and SUSY
  signal events (hashed). Note: the right hand plot is shown for
  illustration purposes only.}

\end{figure}

\begin{figure}[htbp]
\centerline{\epsfxsize=16.0cm \epsfysize=18.0cm \epsfbox{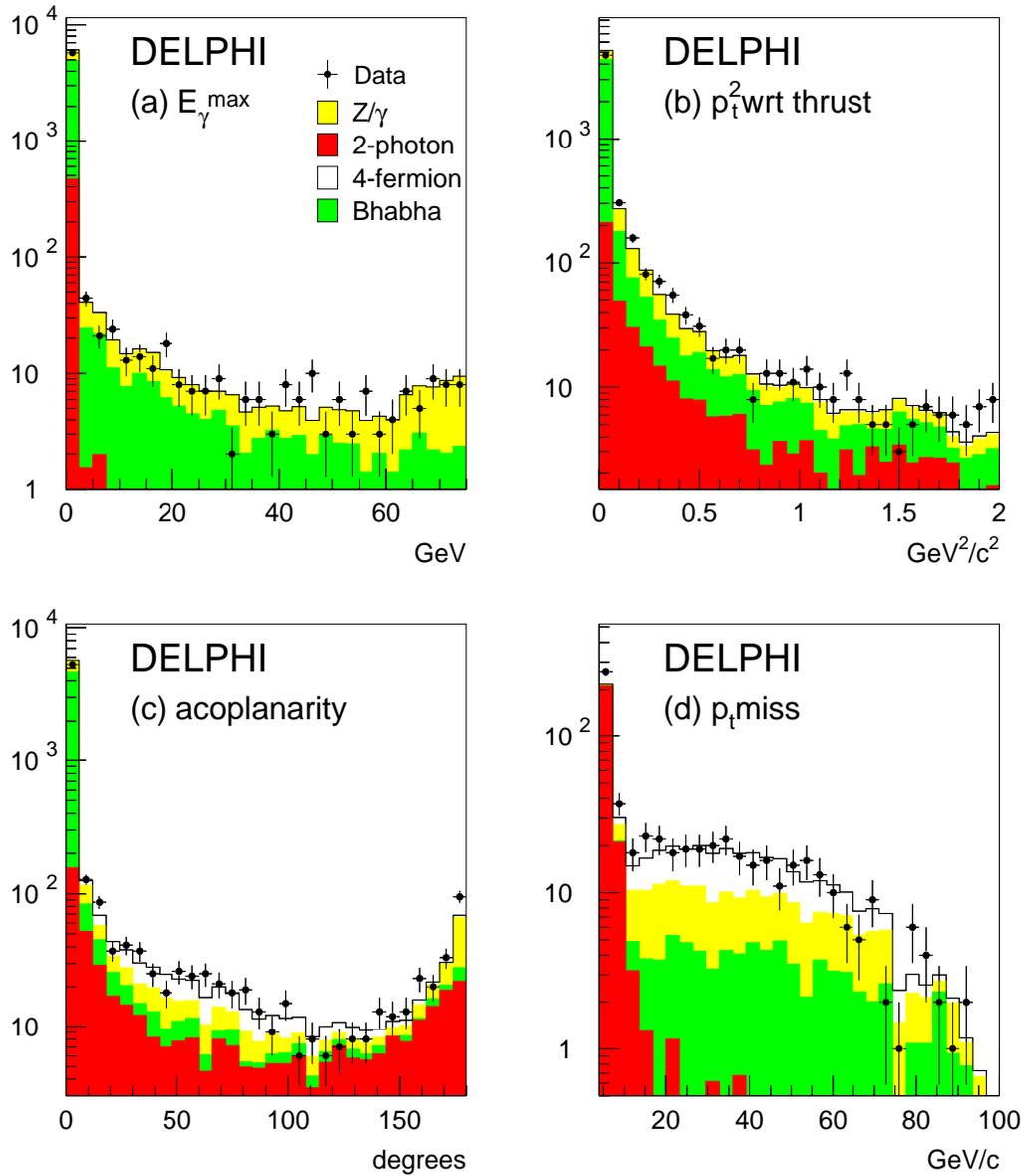}}
\caption{A pre-selection comparison of data and simulated SM events in
  the stau 
  analysis at 189~\GeV. The plots show: (a) Energy of the most energetic,
  isolated photon (b) The square of the transverse momentum with respect to
  the thrust axis (c) The acoplanarity (d) Missing transverse momentum,
  for events with acoplanarity above 10 degrees.
  The dots with error bars show the data, while the simulation is shown
  by the histogram.}  
\label{fig:stau:datmc}
\end{figure}

\begin{figure}[htbp]
\centerline{\epsfysize=18.0cm \epsfbox{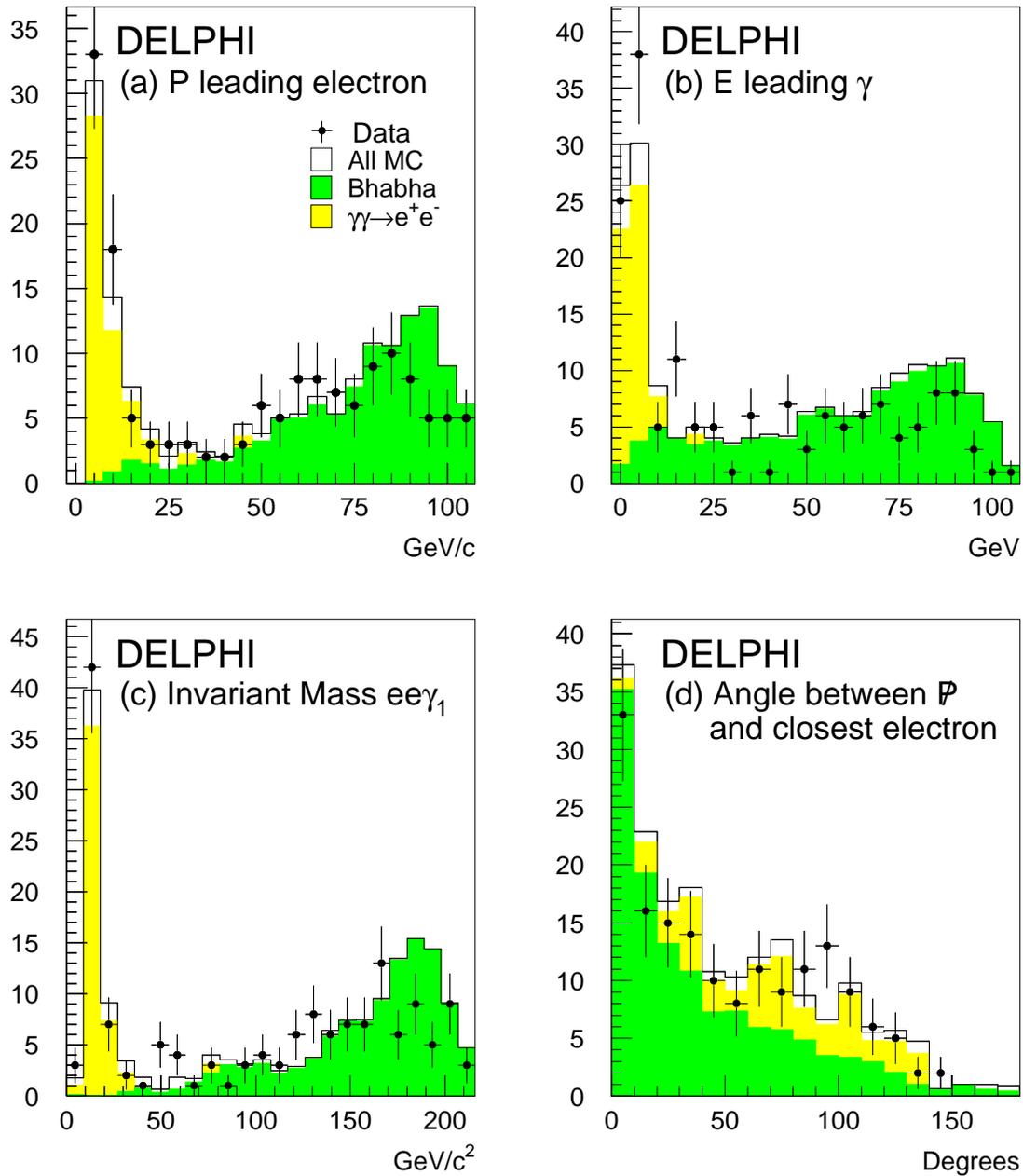}}
\caption{A pre-selection comparison of real data and simulated
  SM events in the electron channel of the 
  cascade decay search. The plots show (a) Momentum of
  leading electron, (b) Energy of leading photon, (c) The invariant mass
  $M_{ee{\gamma}_1}$, (d) The angle between the missing momentum and the
  closest electron.} 
\label{fig:eegg_1}
\end{figure}

\begin{figure}[htbp]
\centerline
\mbox{
  \epsfxsize=7.6cm \epsfysize=7.6cm \epsfbox{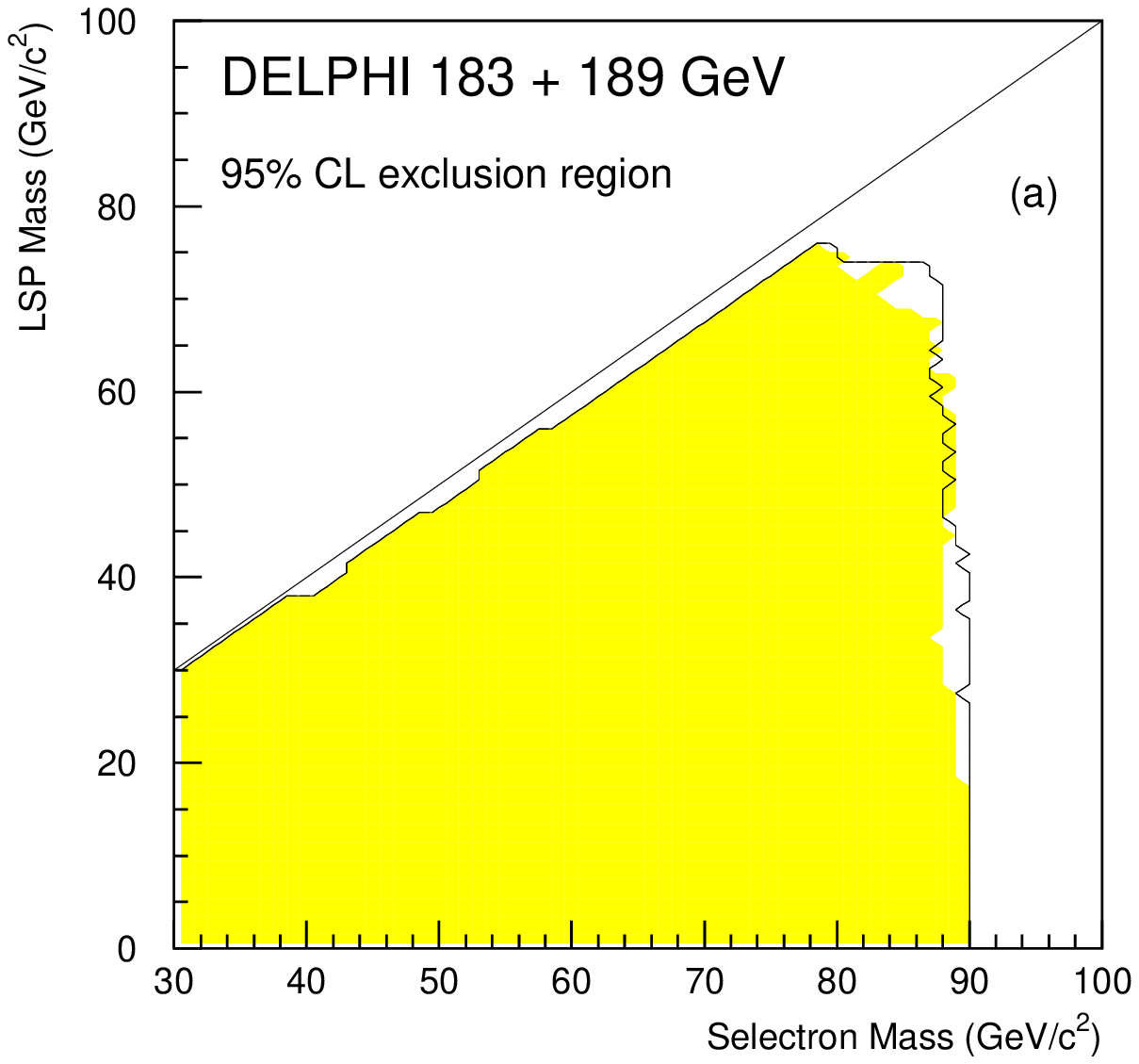}
   \hspace{.1cm}\mbox{\epsfxsize=7.6cm \epsfysize=7.6cm \epsfbox{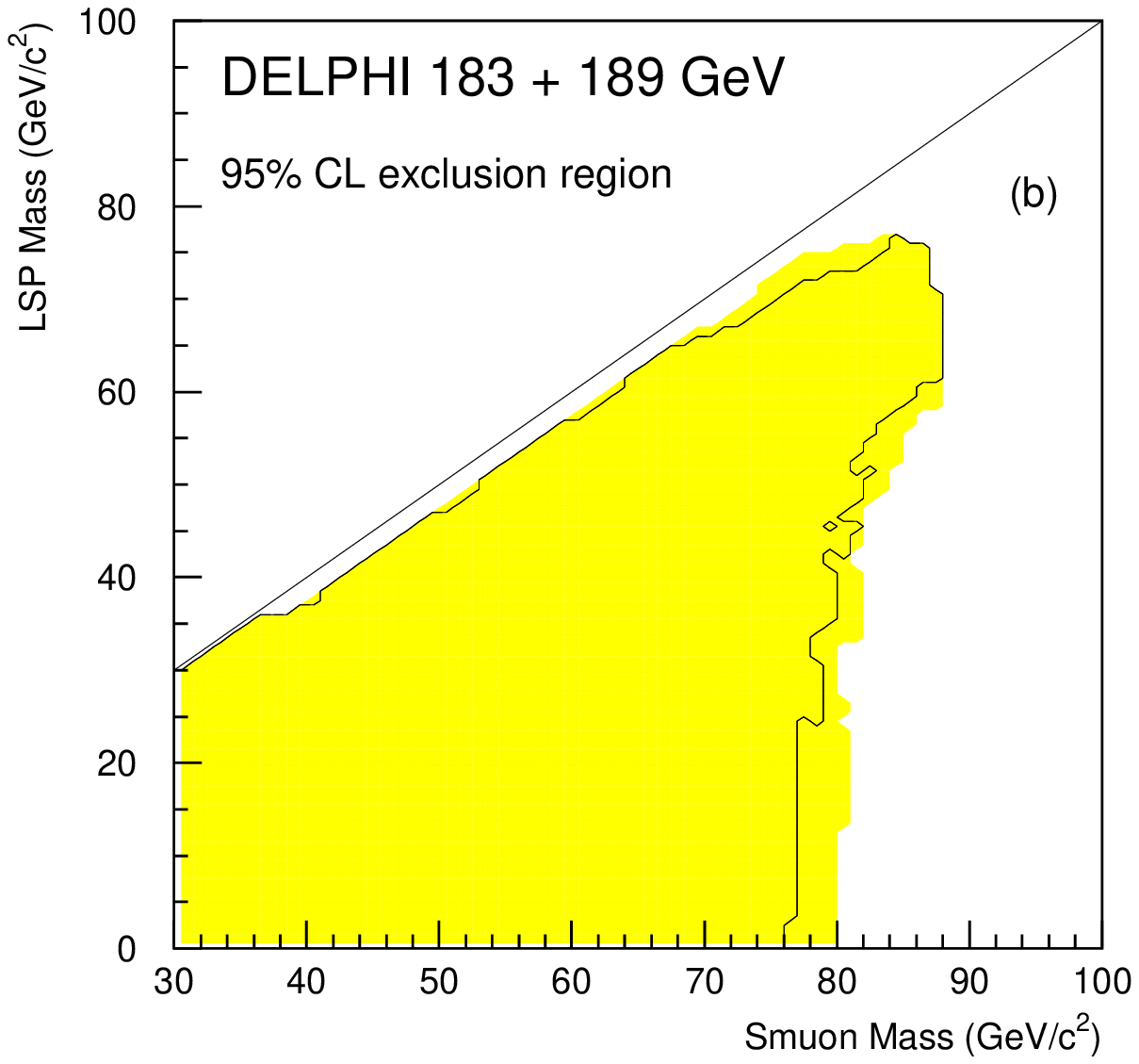}}}

\mbox{
  \hspace{0.2cm}  \epsfxsize=7.6cm \epsfysize=7.6cm \epsfbox{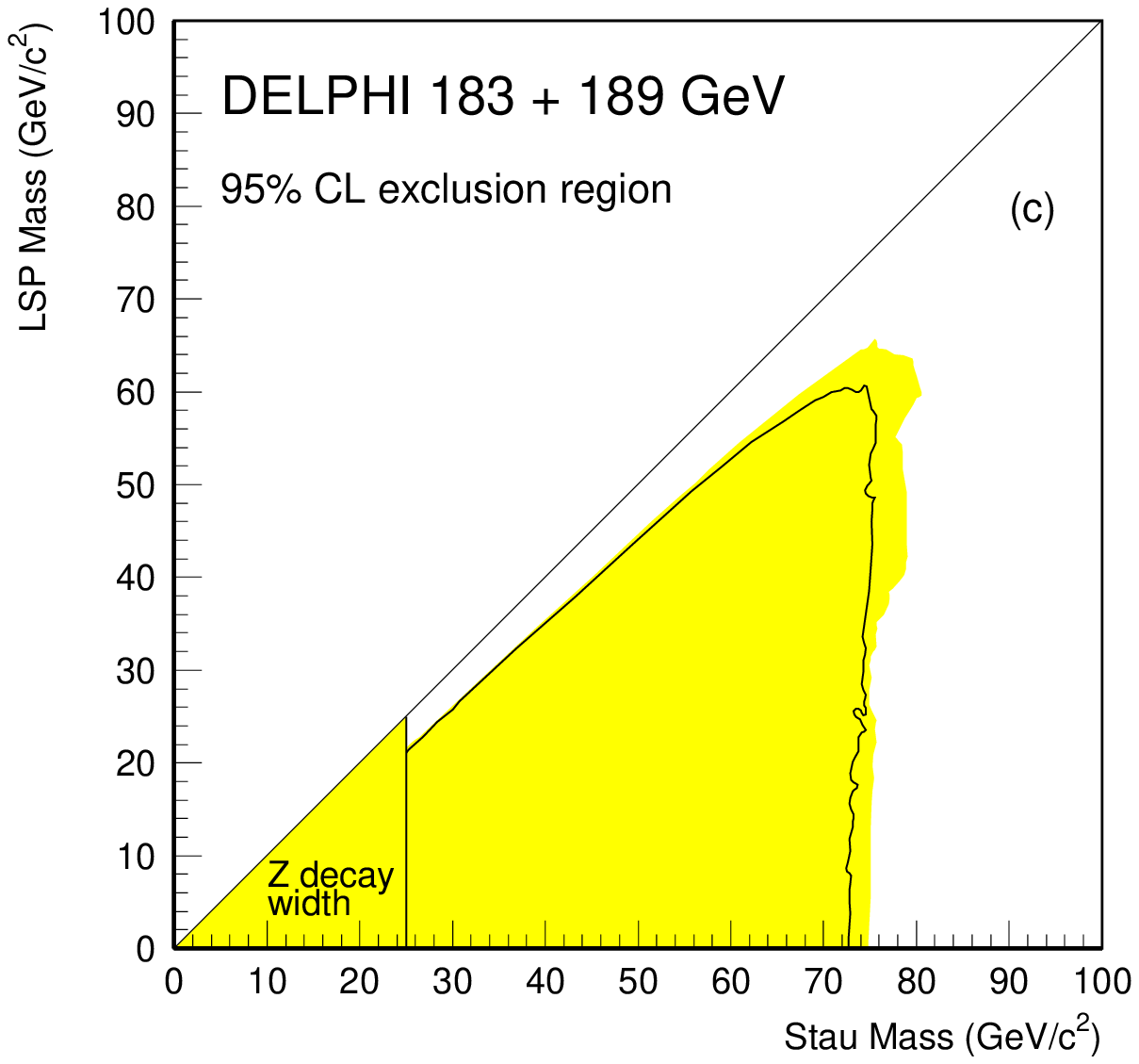}
  \epsfxsize=7.6cm \epsfysize=7.6cm \epsfbox{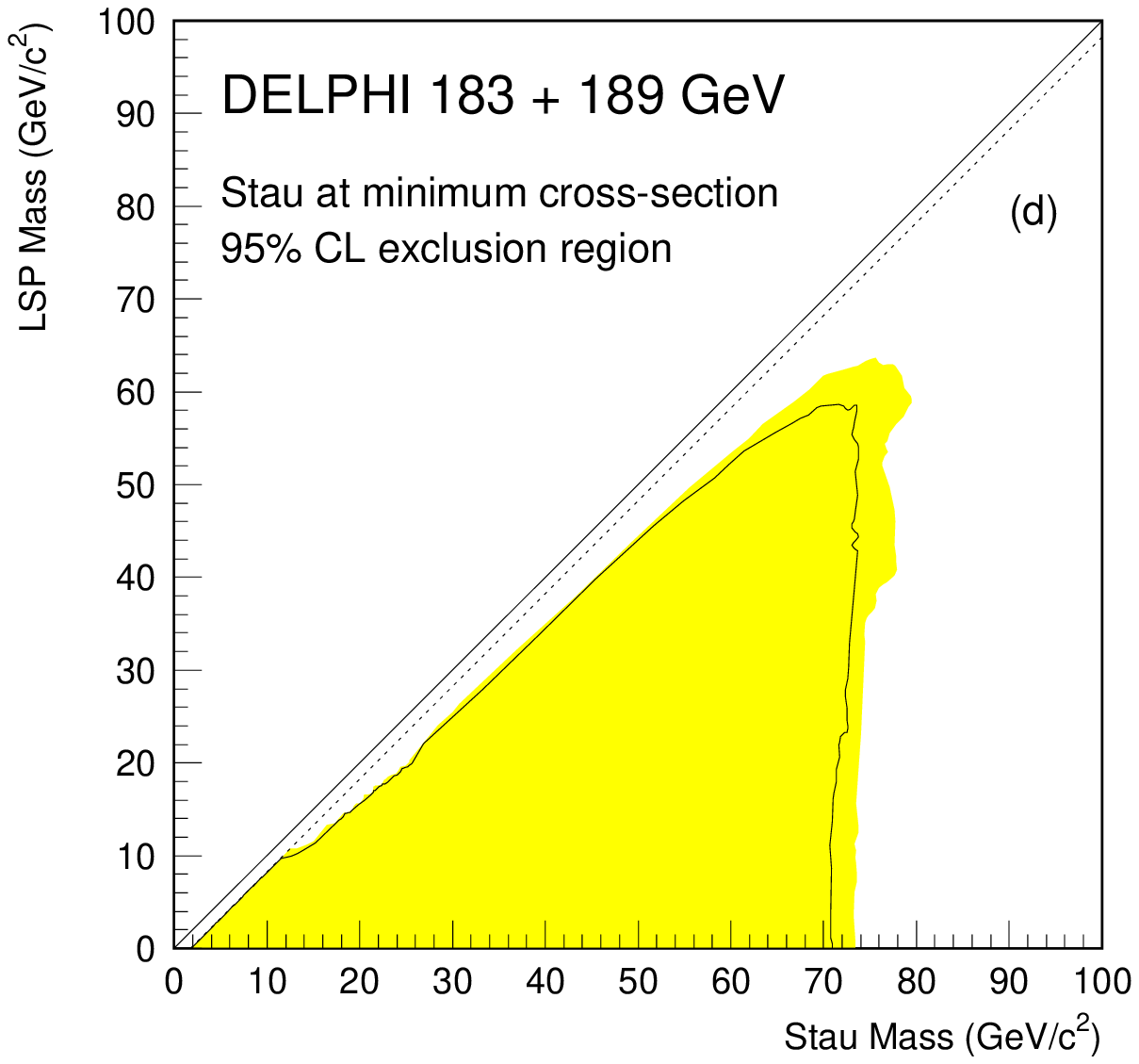}}
\caption{95\% CL exclusion regions
for $\tilde{\ell}\tilde{\ell}$  production in the MSSM.
Figures (a) and (b) show respectively the exclusion region for
$\tilde{e}_R,\tilde{\mu}_R$ 
production in the (${\tilde{\ell}}$,${\lsp}$) mass plane. Figures (c) and (d)
show the mass exclusion regions for the $\tilde{\tau}_R$ and
$\tilde{\tau}_{min}$ in the (${\tilde{\tau}}$,${\lsp}$) mass plane. The
shaded region in the plots shows the obtained exclusion limit, and the solid
line shows the expected limit treating simulated background as data.  
In (d), the dotted line represents $\Delta m$ = $m_{\tau}$. The
limits have been produced using values of tan$\beta$=1.5 and 
$\mu$ = -200~\GeVcc.}  
\label{fig:all:xcl}
\end{figure}

\begin{figure}[!hb]
\begin{center}
\epsfig{file=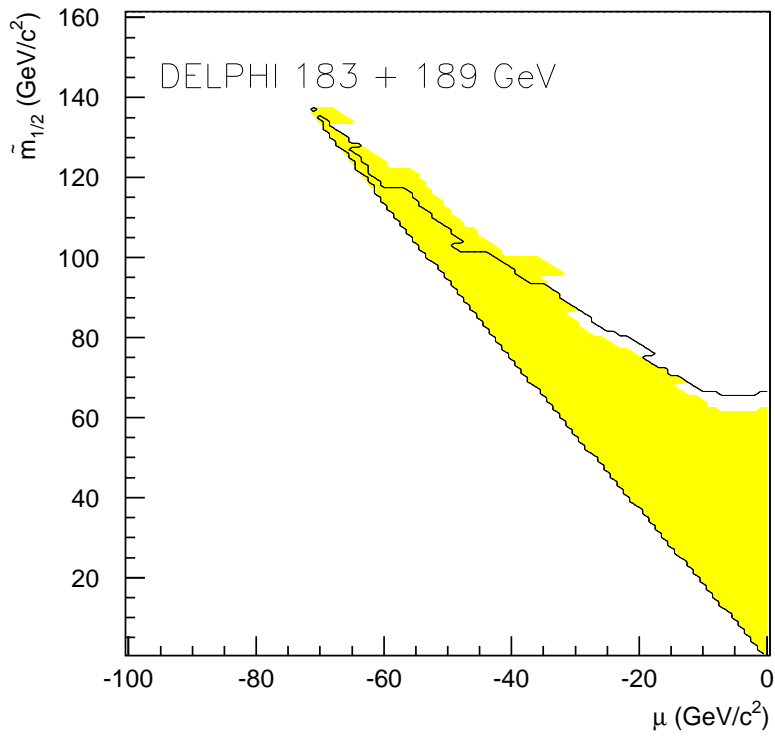,height=120mm}
\caption{Exclusion regions at 95 \% CL in the SUSY parameter space
from the $\ell\ell\gamma\gamma$ events. The shaded region shows the obtained
limit, and the solid line shows the limit treating simulated background as
data. The exclusion region is obtained assuming a slepton mass ,$m_{\tilde
  {\ell}}$ = 80~\GeVcc, and a value of tan$\beta$ = 1.0. The slepton mass
of 80~\GeVcc{} was chosen as it was the highest excluded mass from
the ${\tilde{e}},  {\tilde{\mu}}$ direct decay search.} 
\label{fig:exclusion2}
\end{center}
\end{figure}

\end{document}